\documentclass[10pt, conference]{IEEEtran}
\usepackage{psfrag, amsmath, amssymb, amsfonts, latexsym, eucal, balance, hyperref, indentfirst, graphicx}
\usepackage[tight,footnotesize]{subfigure}
\newcommand{\bm}[1]{\mbox{\boldmath{$#1$}}}

\pdfoutput=1

\DeclareMathAlphabet{\eurm}{U}{eur}{m}{n}
\DeclareMathAlphabet{\mathbsf}{OT1}{cmss}{bx}{n}
\DeclareMathAlphabet{\mathssf}{OT1}{cmss}{m}{sl}
\DeclareMathAlphabet{\mathcsf}{OT1}{cmss}{sbc}{n}

\DeclareSymbolFont{bsfletters}{OT1}{cmss}{bx}{n}  
\DeclareSymbolFont{ssfletters}{OT1}{cmss}{m}{n}
\DeclareMathSymbol{\bsfGamma}{0}{bsfletters}{'000}
\DeclareMathSymbol{\ssfGamma}{0}{ssfletters}{'000}
\DeclareMathSymbol{\bsfDelta}{0}{bsfletters}{'001}
\DeclareMathSymbol{\ssfDelta}{0}{ssfletters}{'001}
\DeclareMathSymbol{\bsfTheta}{0}{bsfletters}{'002}
\DeclareMathSymbol{\ssfTheta}{0}{ssfletters}{'002}
\DeclareMathSymbol{\bsfLambda}{0}{bsfletters}{'003}
\DeclareMathSymbol{\ssfLambda}{0}{ssfletters}{'003}
\DeclareMathSymbol{\bsfXi}{0}{bsfletters}{'004}
\DeclareMathSymbol{\ssfXi}{0}{ssfletters}{'004}
\DeclareMathSymbol{\bsfPi}{0}{bsfletters}{'005}
\DeclareMathSymbol{\ssfPi}{0}{ssfletters}{'005}
\DeclareMathSymbol{\bsfSigma}{0}{bsfletters}{'006}
\DeclareMathSymbol{\ssfSigma}{0}{ssfletters}{'006}
\DeclareMathSymbol{\bsfUpsilon}{0}{bsfletters}{'007}
\DeclareMathSymbol{\ssfUpsilon}{0}{ssfletters}{'007}
\DeclareMathSymbol{\bsfPhi}{0}{bsfletters}{'010}
\DeclareMathSymbol{\ssfPhi}{0}{ssfletters}{'010}
\DeclareMathSymbol{\bsfPsi}{0}{bsfletters}{'011}
\DeclareMathSymbol{\ssfPsi}{0}{ssfletters}{'011}
\DeclareMathSymbol{\bsfOmega}{0}{bsfletters}{'012}
\DeclareMathSymbol{\ssfOmega}{0}{ssfletters}{'012}

\begin{document}

\title{Separating the Wheat from the Chaff:\\ Sensing Wireless Microphones in TVWS}
\author{
\authorblockN{Huanhuan Sun, Taotao Zhang, Wenyi Zhang}
\authorblockA{
Department of Electronic Engineering and Information Science\\
University of Science and Technology of China, Hefei, 230027, China\\
Email: wenyizha@ustc.edu.cn\\
\footnotesize{* H. Sun and T. Zhang contributed equally to this work.}}
}
\maketitle

\begin{abstract}
This paper summarizes our attempts to establish a systematic approach that overcomes a key difficulty in sensing wireless microphone signals, namely, the inability for most existing detection methods to effectively distinguish between a wireless microphone signal and a sinusoidal continuous wave (CW). Such an inability has led to an excessively high false alarm rate and thus severely limited the utility of sensing-based cognitive transmission in the TV white space (TVWS) spectrum. Having recognized the root of the difficulty, we propose two potential solutions. The first solution focuses on the periodogram as an estimate of the power spectral density (PSD), utilizing the property that a CW has a line spectral component while a wireless microphone signal has a slightly dispersed PSD. In that approach, we formulate the resulting decision model as an one-sided test for Gaussian vectors, based on Kullback-Leibler distance type of decision statistics. The second solution goes beyond the PSD and looks into the spectral correlation function (SCF), proposing an augmented SCF that is capable of revealing more features in the cycle frequency domain compared with the conventional SCF. Thus the augmented SCF exhibits the key difference between CW and wireless microphone signals. Both simulation results and experimental validation results indicate that the two proposed solutions are promising for sensing wireless microphones in TVWS.
\end{abstract}

\section{Introduction}
\label{sec:intro}

The digital TV transition has created the opportunity of realizing cognitive radio in the TV white space (TVWS) spectrum \cite{fcc10:order}-\cite{shellhammer11:chapter}. In the envisioned TVWS systems, TV band devices (TVBDs) cognitively operate in TV spectrum that is not used by TV broadcasting services or certain low-power auxiliary stations such as wireless microphones \cite{fcc47cfr}.

Spectrum sensing is a functionality for TVBDs, via signal processing techniques, to identify the presence or absence of TV or wireless microphone signals with high reliability, thus providing an effective protection mechanism for those incumbent services. While sensing for TV signals such as ATSC/NTSC/DVB-T has been demonstrated as feasible \cite{oet08:report}-\cite{balamurthi11:dyspan}, sensing for wireless microphone signals remains a challenging problem. Most tested sensing prototypes have not exhibited satisfactory performance, especially in realistic environment \cite{oet08:report}.

There have been a number of works \cite{han06:icact}-\cite{zhang10:globecom} aiming at improving the accuracy of wireless microphone sensing. Among those works, despite of their different methodologies, a common ingredient is that the background noise is modeled as Gaussian, either white, or colored in the presence of adjacent channel interference. Since wireless microphone signal is a narrowband frequency-modulated (FM) waveform \cite{fcc47cfr}, its power spectral density (PSD) or (the magnitude of) spectral correlation function (SCF) manifests itself like one or several peaks in the perceptually ``smooth'' noise background. Visually, such a situation is like ``telling the tree from the grassland''.

\begin{figure}[!t]
\centering
\includegraphics[width=3in]{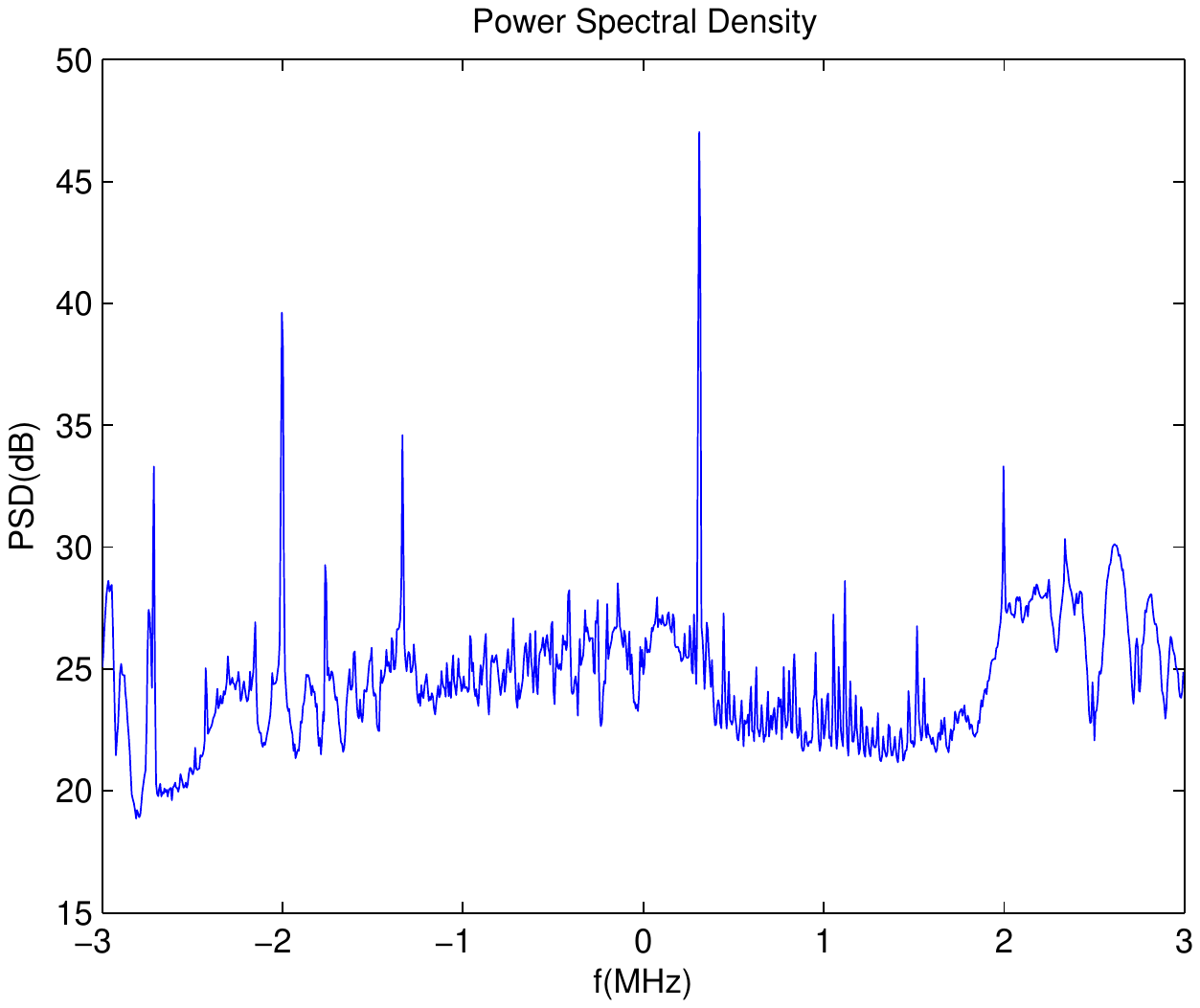}
\caption{\cite[Fig. 7]{shellhammer09:ita} Received periodogram of a typical TV channel with wireless microphone signal and narrowband interference signals. The channel center frequency is converted to the baseband, and the scale of the y-axis is not calibrated. The wireless microphone signal is at $f = -2$MHz, and the other spikes are from unknown emissions.}
\label{fig:wm-spurs}
\end{figure}

In realistic environment, however, the perhaps unpleasant reality is that the background noise is seldom pure Gaussian. As is well known, narrowband interference such as spurious emissions, unintentional transmissions, leakage, and inter-modulation are inevitable and universal in electronic devices. Available regulatory rules do strictly limit their interference upon communications. For spectrum sensing which is supposed to be capable of detecting weak signals well below the noise floor, however, the impact of those narrowband interference sources becomes crucial. Figure \ref{fig:wm-spurs} illustrates such a situation.

In retrospect, the main difficulty with wireless microphone sensing has been largely related to such narrowband interference, which ``fools'' sensing algorithms to erroneously report that an available TVWS is being used, thus incurring excessively high false alarm rate and severely reducing the amount of sensed TVWS spectrum. This difficulty has been realized by researchers mainly from the industry side \cite{shellhammer09:ita}\cite{802fcc}\cite{sharkey09:petition}. The foremost task, therefore, is to distinguish between an FM wireless microphone signal and a sinusoidal continuous wave (CW) which models the narrowband interference, and the situation thus is more like ``telling the wheat from the chaff'', in that it is necessary to extract fine features of the received signal in order to make the correct decision.

Recently, the sensing prototype reported in \cite{balamurthi11:dyspan} used an ad hoc algorithm to decide whether a received narrowband signal is from a wireless microphone or is a CW from an interference source. The algorithm computes certain features of the periodogram of the received signal, and tests those features against rules that have been fine-tuned via training several specific makes of wireless microphones. In the current paper, we attempt to establish a more systematic understanding of the underlying decision problem, and upon such a understanding, develop more systematic sensing methods, that are potentially suited for general scenarios without extensive training and fine-tuning. Specifically, we focus on two approaches that are based on the periodogram and an augmented SCF, respectively. The first approach focuses on the periodogram, which is an estimate of the PSD, utilizing the property that a CW has a line spectral component while a wireless microphone signal, as FM, has a slightly dispersed PSD. We formulate the resulting decision model as an one-sided hypothesis test for Gaussian vectors, and develop the decision statistic based on the Kullback-Leibler distance between the empirical distribution of the observed periodogram and the distribution of the periodogram under CW hypothesis. The second approach goes beyond the PSD and looks into the SCF. More specifically, we propose to use an augmented SCF that combines the conventional SCF and a proposed conjugate SCF. The augmented SCF is capable of revealing more features in the cycle frequency domain compared with the conventional SCF, and thus exhibits the key difference between CW and wireless microphone signals. As both simulation results and experimental validation results indicate, the two proposed approaches are promising solutions for sensing wireless microphones in TVWS.

Some remarks on the relevance of the topic are in order. The Federal Communication Commission (FCC) of the United States recently has waived the mandatory requirements on spectrum sensing, turning to geolocation database access as the main means of primary service protection for TVWS systems \cite{fcc10:order}. However, this change in the policy should not be deemed as the end of TVWS spectrum sensing. First, the FCC rules still encourage further research into spectrum sensing and keep the option of ``sensing only'' TVBDs, which sometimes may be more flexible to use than TVBDs that use only geolocation database access. The sensing threshold for wireless microphone signals in the FCC rules is set as $-107$dBm, which, when considering a $6$MHz TV channel of background thermal noise only and a typical handset RF circuitry implementation \cite{balamurthi11:dyspan}, would correspond to a signal-to-noise ratio (SNR) lower than $-10$dB. Second, in a deployed geolocation database service, it will be useful or even necessary for the administrators to have some TVBDs or some surveillance nodes possess the sensing capability, so that they can, when needed, verify the validity of records in the database. Third, the understanding we develop herein for wireless microphone sensing may prove useful under other scenarios, say, cognitive radio in TVWS of other countries, or in other spectrum bands for cognitive usage.

The remaining part of this paper is organized as follows. Section \ref{sec:formulation} formulates the decision-theoretic model of wireless microphone sensing. Sections \ref{sec:psd} and \ref{sec:scf} respectively develop the sensing approaches based on the periodogram and an augmented SCF of the received signals, describe the sensing methods, and present their performance by simulations. Section \ref{sec:experiment} presents the experimental validation of the two approaches. Section \ref{sec:conclusion} concludes the paper.

\section{Problem Formulation}
\label{sec:formulation}

Since a wireless microphone may choose its carrier frequency rather flexibly, upon receiving the signals from a TV channel, the first task is to identify the frequency location(s) of potential wireless microphone signal(s). To accomplish this task, a simple and effective method (see, e.g., \cite{ghosh08crowncom}\cite{balamurthi11:dyspan}) is to compute the periodogram of the signal, scan through the periodogram to search for peaks that exceed a certain prescribed threshold. In order to focus our attention on the core problem, we assume that the frequency scan has been done successfully, and that for each thus identified spectral peak, the resulting binary hypothesis testing problem is
\begin{eqnarray}
\label{eqn:ct-hypo-1}
&&\mathcal{H}_1\; \mbox{(wireless microphone)}:\nonumber\\
&& \hspace{-0.5in}x(t) = A \cos\left(2\pi f_c t + 2\pi \beta \int_0^t m(\tau)d\tau + \varphi\right) + w(t),\\
\label{eqn:ct-hypo-0}
&&\mathcal{H}_0\; \mbox{(CW)}:\nonumber\\
&& \hspace{-0.5in}x(t) = A \cos(2\pi f_c t + \varphi) + w(t).
\end{eqnarray}
In the problem formulation, $A$ denotes the carrier magnitude, $f_c$ denotes the carrier frequency, and $\varphi$ denotes the signal phase uncertainty, which may be deemed as uniformly distributed in $[0, 2\pi)$. For the FM wireless microphone signal, $m(t)$ represents the message which may be the voice of the speaker, and $\beta$ is the frequency deviation, whose value determines the degree of dispersion of the resulting FM spectrum. The noise $w(t)$ is modeled as zero-mean white Gaussian, and in this paper we do not consider the issue of colored noise due to adjacent channel leakage. Inspecting the problem formulation, we clearly see that its key distinction from the existing works is that the null hypothesis $\mathcal{H}_0$ is no longer pure Gaussian noise, but contains a CW component, which closely captures the ambiguity due to the narrowband interference encountered in realistic TV channels \cite{shellhammer09:ita}\cite{sharkey09:petition}\cite{balamurthi11:dyspan}.

Since the message $m(t)$ is a random process, the maximum-likelihood (ML) detector would be of an estimator-correlator architecture, which first forms the minimum-mean-squared-error (MMSE) estimate of $A \cos\left(2\pi f_c t + 2\pi \beta \int_0^t m(\tau)d\tau + \varphi\right)$ as if the actual hypothesis is $\mathcal{H}_1$, and then treats the estimated $\hat{m}(t)$ as the actual message $m(t)$ to compute the likelihood ratio between the two hypotheses \cite{kailath98:it}. In reality, however, since the prior statistical knowledge of $m(t)$ is sophisticated and usually unavailable, and the signal part is typically weak compared with the noise, it is hardly feasible to implement the ML detector with tolerable complexity. Therefore, in this work, we turn to suboptimal approaches to the problem.

In order to process the received signal digitally, we need to down-convert, low-pass filter, and sample $x(t)$. Denote by $f_s$ the sampling rate for low-pass filtered baseband signal, and let $T_s = 1/f_s$. We have the discrete-time version of the hypothesis testing problem (\ref{eqn:ct-hypo-1}) (\ref{eqn:ct-hypo-0}) as
\begin{eqnarray}
\label{eqn:dt-hypo-1}
&&\mathcal{H}_1\; \mbox{(wireless microphone)}:\nonumber\\
&&\quad\quad \hspace{-0.5in}x[n] = \frac{A}{2} e^{j \left(2\pi \beta \int_0^{n T_s} m(\tau)d\tau + \varphi\right)} + w[n],\\
\label{eqn:dt-hypo-0}
&&\mathcal{H}_0\; \mbox{(CW)}:\nonumber\\
&&\quad\quad \hspace{-0.5in}x[n] = \frac{A}{2} e^{j \varphi} + w[n].
\end{eqnarray}
Herein $w[\cdot]$ represents the bandlimited white circularly-symmetric complex Gaussian noise, with variance denoted by $\sigma^2$. The SNR is therefore $\mathrm{SNR} = A^2/(4\sigma^2)$.

{\it Remark 1:} In this paper, we do not consider the impact of multipath fading on sensing. On one hand, even without multipath fading, the hypothesis testing problem (\ref{eqn:dt-hypo-1}) (\ref{eqn:dt-hypo-0}) itself is already a new challenge beyond what has been considered in the literature, and thus we choose to focus on the core problem at this stage. On the other hand, since both an FM wireless microphone signal and a CW are rather narrowband, their spectral shapes are unlikely to be noticeably distorted by channel multipath, except for a frequency-flat attenuation, which effectively acts as a scaling on the SNR.

{\it Remark 2:} A noteworthy fact is that, in realistic systems, since the carrier frequency $f_c$ is typically estimated by scanning through the periodogram, the finite frequency resolution inevitably leads to a certain nonzero frequency offset between $f_c$ and the estimated $\hat{f}_c$. The magnitude of the frequency offset is a number between zero and the frequency resolution, and it will ``fatten'' the resulting periodogram of the signal (wireless microphone signal or CW). As will be made clear in Section \ref{sec:psd}, the ``fattening'' effect would increase the likelihood of false alarm, and thus it is desirable to eliminate the frequency offset at least when the underlying hypothesis is $\mathcal{H}_0$. In Appendix \ref{app:off-est} we describe one such method, which has proven effective in both simulation and experiments.

\section{Periodogram-based Solution}
\label{sec:psd}

\subsection{Rationale of Using Periodogram and One-sided Test}

The periodogram is an estimate of the PSD of the received signal embedded in noise. The starting point of using periodogram here is a well-known fact that, CW has a line PSD whereas a wireless microphone signal has a slightly dispersed PSD due to the modulation by the message $m(t)$. So extracting fine features of the periodogram will enable the detector to distinguish between CW and wireless microphone.

The approach we take in this work, unlike the ad hoc approach in \cite{balamurthi11:dyspan} where ``behavioral'' features like main lobe width are extracted from the periodogram to form the decision statistics, treats the periodogram as a vector of Gaussian random variables and establishes the decision rule based on such a Gaussian model. Concretely, for $M$ non-overlapping length-$N$ segments of the received signals, we compute their averaged periodogram, denoted by $\bm{\xi}_a = (\xi_a[0],\; \xi_a[1],\; \ldots, \xi_a[N - 1])$, as the estimate of the PSD. Due to the central limit theorem, for moderate or large $M$ we may approximate $\bm{\xi}_a$ as a vector of Gaussian random variables. Hence if we could obtain the means and covariances of $\bm{\xi}_a$ under both $\mathcal{H}_0$ and $\mathcal{H}_1$, we could directly utilize the likelihood ratio test to decide whether a received signal is a CW or a wireless microphone signal.

A realistic difficulty with likelihood ratio tests, unfortunately, is that the mean and covariance of $\bm{\xi}_a$ is generally unknown under $\mathcal{H}_1$, due to the lack of prior knowledge on the FM implementation details of wireless microphones. Different makes of wireless microphones may have quite different implementations. Their choices of the frequency deviation $\beta$ differ and the statistics of $m(t)$ highly depend on the message characteristics and the audio processing methods. This difficulty thus motivates our idea of using one-sided test. That is, we only specify the statistics of $\bm{\xi}_a$ under the null hypothesis $\mathcal{H}_0$, and the one-sided test aims at telling whether a received signal is from $\mathcal{H}_0$ or not. Whenever it is decided that the null hypothesis is to be rejected, we interpret the rejection as an indication of the occurrence of wireless microphone signals.

There are numerous one-sided test rules, and in this work, we follow the information-theoretic approach in \cite{kullback59:book}, using the Kullback-Leibler (KL) distance between the empirical distribution of the averaged periodogram and the actual distribution of $\bm{\xi}_a$ under $\mathcal{H}_0$. The idea is that since the KL distance vanishes only if its two distributions coincide, we would accept the null hypothesis only when the KL distance is sufficiently small.

A side note is that one may also use other, perhaps more refined, PSD estimates to replace periodogram in computing the statistics. We, however, do not expect that refinement to yield significant performance improvement. There are two considerations. First, the lack of prior statistical knowledge for wireless microphone signals prevents the usage of many data-dependent, parametric PSD estimation methods. Second, most nonparametric PSD estimation methods tend to ``fatten'' the main lobe of signals, thus rendering it more difficult to distinguish CW from wireless microphone signals.

\subsection{Method Description}

Regarding the signal model of CW in (\ref{eqn:dt-hypo-0}), in Appendix \ref{app:off-est} we evaluate the means and variances of its periodogram. Specifically, since the frequency offset is eliminated and $M$ periodograms are averaged, we have from (\ref{mean_X}) and (\ref{var_X}) that
\begin{eqnarray}
\mu[k] \triangleq \mathbb{E}\left\{\xi_a[k]\right\} &=& \sigma^2 (N \cdot \mathrm{SNR} + 1) \quad \mbox{if}\; k = 0\nonumber\\
&=& \sigma^2 \quad \mbox{otherwise}\\
\eta[k, k] \triangleq \mathrm{var}\left\{\xi_a[k]\right\} &=& \frac{\sigma^4}{M} (2N \cdot \mathrm{SNR} + 1) \quad \mbox{if}\; k = 0\nonumber\\
&=& \frac{\sigma^4}{M} \quad \mbox{otherwise}.
\end{eqnarray}
Furthermore, without frequency offset, we can directly verify that the off-diagonal elements $\{\eta[k, l]\}_{k \neq l}$ in the covariance matrix of $\bm{\xi}_a$ are all zero. Denote the mean vector of $\bm{\xi}_a$ by $\bm{\mu} = (\mu[0],\; \mu[1],\;\ldots, \mu[N - 1])$, and the covariance matrix of $\bm{\xi}_a$ by
\begin{eqnarray*}
\bm{\eta} = \left[\begin{array}{cccc}
  \eta[0, 0] & 0 & \cdots & 0\\
  0 & \eta[1, 1] & \cdots & 0\\
  \vdots & \vdots & \ddots & \vdots\\
  0 & 0 & \cdots & \eta[N - 1, N - 1]
\end{array}\right].
\end{eqnarray*}
So by our approximation, $\bm{\xi}_a \sim \mathcal{N}\left(\bm{\mu}, \bm{\eta}\right)$.

On the other hand, upon computing the averaged (over $M$ non-overlapping length-$N$ segments) periodogram of the received signal, which may be either a CW or a wireless microphone signal, we can evaluate its empirical mean and covariance. Denote by $\bm{\xi}_e$ the empirical averaged periodogram, and by $\bm{\mu}_e$ and $\bm{\eta}_e$ its empirical mean vector and covariance matrix respectively. According to the one-sided decision rule using KL distance \cite{kullback59:book}, the decision statistic is given by
\begin{eqnarray}
T_{p, 0} &\triangleq& \mathcal{D}\left(\mathcal{N}(\bm{\mu}_e, \bm{\eta}_e) \| \mathcal{N}(\bm{\mu}, \bm{\eta})\right) \nonumber\\
&=& \frac{1}{2} (\bm{\mu}_e - \bm{\mu})^T \bm{\eta}^{-1} (\bm{\mu}_e - \bm{\mu}) +\nonumber\\
 &&\quad \frac{1}{2} \left[\mathrm{trace}(\bm{\eta}^{-1} \bm{\eta}_e - \bm{I}) + \log \frac{\det \bm{\eta}}{\det \bm{\eta}_e}\right].
\end{eqnarray}

In $T_p$ we need to compute the empirical covariance matrix $\bm{\eta}_e$ and use all the $N$ elements of $\bm{\xi}$. Both of these practices are computation-intensive and usually sensitive to outliers in real data. Therefore, in order to simplify the detector implementation and to improve its immunity to outliers, we use in this work the following decision statistic,
\begin{eqnarray}
\label{eqn:Tp}
T_p \triangleq \frac{1}{2} (\bm{\mu}_{e, L} - \bm{\mu}_L)^T \bm{\eta}_L^{-1} (\bm{\mu}_{e, L} - \bm{\mu}_L),
\end{eqnarray}
where $L \leq N$ is a window size,\\
$\bm{\mu}_L = (\mu[0],\; \mu[1],\; \ldots, \mu[L - 1])$,\\
$\bm{\mu}_{e, L} = (\mu_e[0],\; \mu_e[1],\; \ldots, \mu_e[L - 1])$,\\
and $\bm{\eta}_L$ is the submatrix of $\bm{\eta}$ formed by its first $L$ rows and first $L$ columns.

Given a threshold $\gamma$, the decision rule is as follows,
\begin{eqnarray*}
\hat{\mathcal{H}} = \mathcal{H}_0 \quad\mbox{if}\; T_p \leq \gamma;\; \mbox{and}\; \mathcal{H}_1 \quad\mbox{otherwise}.
\end{eqnarray*}

By adjusting $\gamma$ we control the false alarm rate. This needs to be accomplished by the aid of Monte Carlo simulations. A first-cut approximation may be used for not too small target false alarm rates, based on the Gaussian approximation of $\bm{\xi}_a$. That is, from the decision statistic in (\ref{eqn:Tp}), when the Gaussian approximation is well-satisfied, we have $T_p \sim \chi^2(L)$.

\subsection{Simulation Results}
\label{subsec:psd-simu}

We use Monte Carlo simulation to verify our periodogram-based sensing method. We generate bandpass CW and FM wireless microphone signals contaminated by white Gaussian noise, then down-convert, low-pass filter, and down-sample the signals to obtain the baseband discrete-time signals as in (\ref{eqn:dt-hypo-1}) and (\ref{eqn:dt-hypo-0}). The SNR is set with respect to a $8$MHz bandwidth (the TV channels in China are $8$MHz), and varies between $-15$dB and $-23$dB. For simplicity and generality, when generating FM wireless microphone signals, we let the message $m(t)$ be a Brownian motion. For making decision, fifteen segments of signals each of time duration $10$ms are used to generate the decision statistics.

In our simulation, we mainly aim at revealing the impact of the following three key parameters:
\begin{itemize}
\item SNR
\item $L$: the window size when computing $T_p$
\item $\beta$: the frequency deviation in FM wireless microphone signals
\end{itemize}

\begin{figure}[!t]
\centering
\includegraphics[width=3.3in]{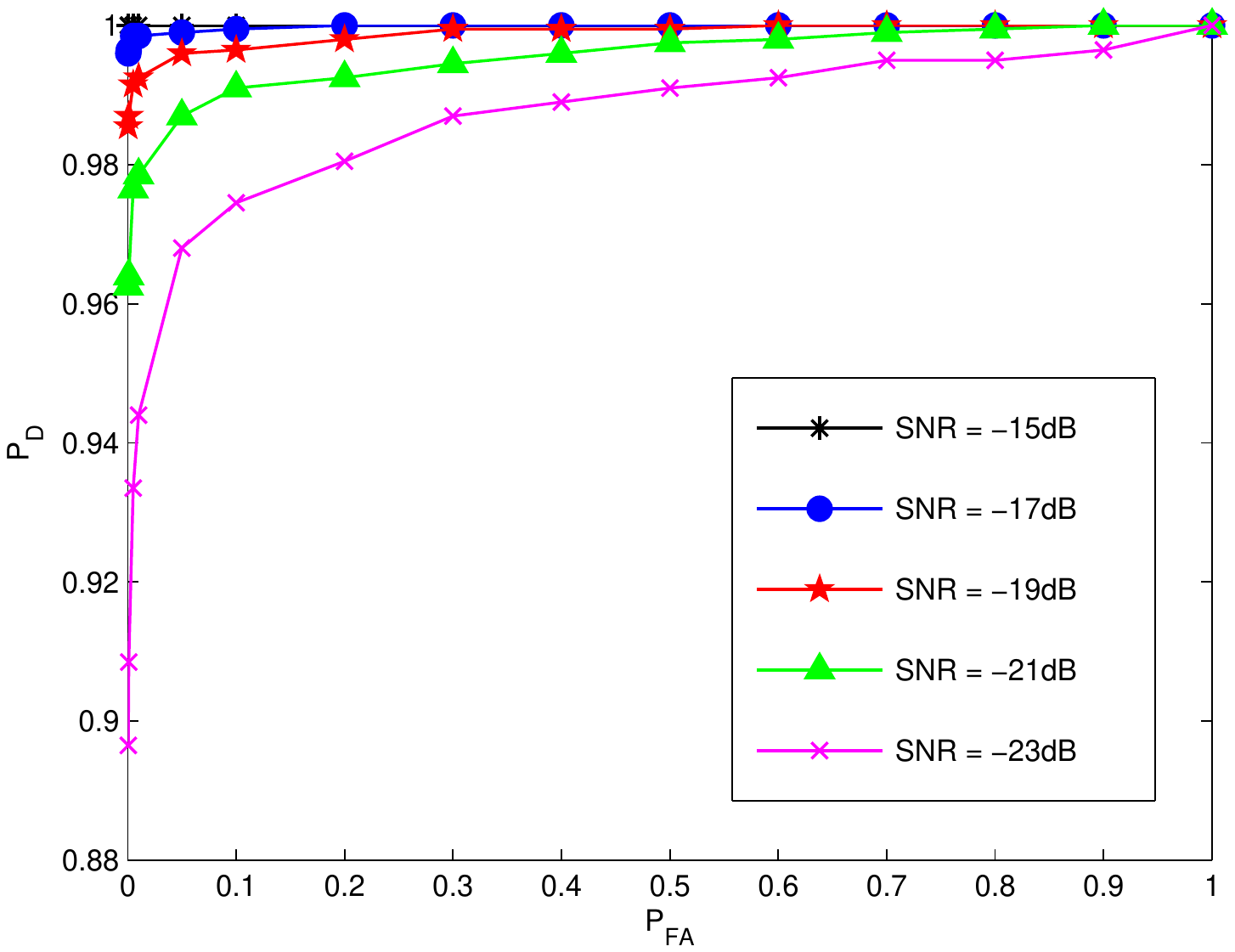}
\caption{ROC curves of periodogram-based sensing method under different SNRs.}
\label{fig:snr_diff}
\end{figure}

Figure \ref{fig:snr_diff} displays the receive operating characteristic (ROC) curves of the detection method, for $L = 11$, $\beta = 0.7$, under different values of $\mathrm{SNR}$. The x-axis is the false alarm rate and the y-axis is the corresponding detection rate (or, one minus the miss rate). It is clearly seen that increasing the SNR dramatically improves the detector performance. When the SNR is (or higher than) $-21$dB, the detection rate exceeds $98\%$ with the false alarm rate being $5\%$.

\begin{figure}[!t]
\centering
\includegraphics[width=3.3in]{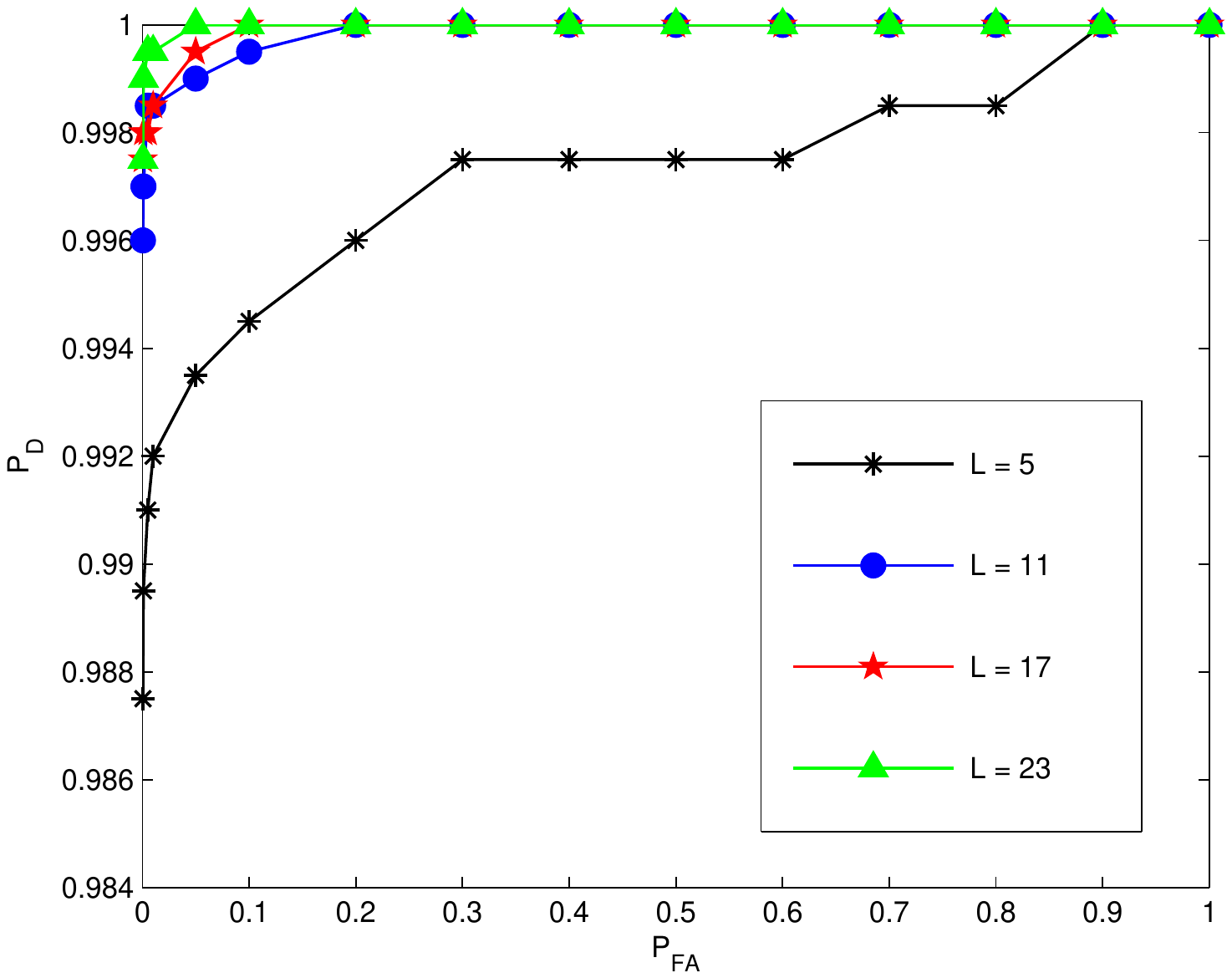}
\caption{ROC curves of periodogram-based sensing method under different $L$s.}
\label{fig:L_diff}
\end{figure}

Figure \ref{fig:L_diff} displays the ROC curves for $\mathrm{SNR} = -17$dB, $\beta = 0.7$, under different values of $L$. It is seen that the detector performance improves with $L$, thus indicating the tradeoff between performance and complexity. Furthermore, increasing $L$ beyond $11$ does not appear to yield substantial performance gain. For $L = 5$, the detection rate exceeds $98\%$ with the false alarm rate being as low as $1\%$.

\begin{figure}[!t]
\centering
\includegraphics[width=3.3in]{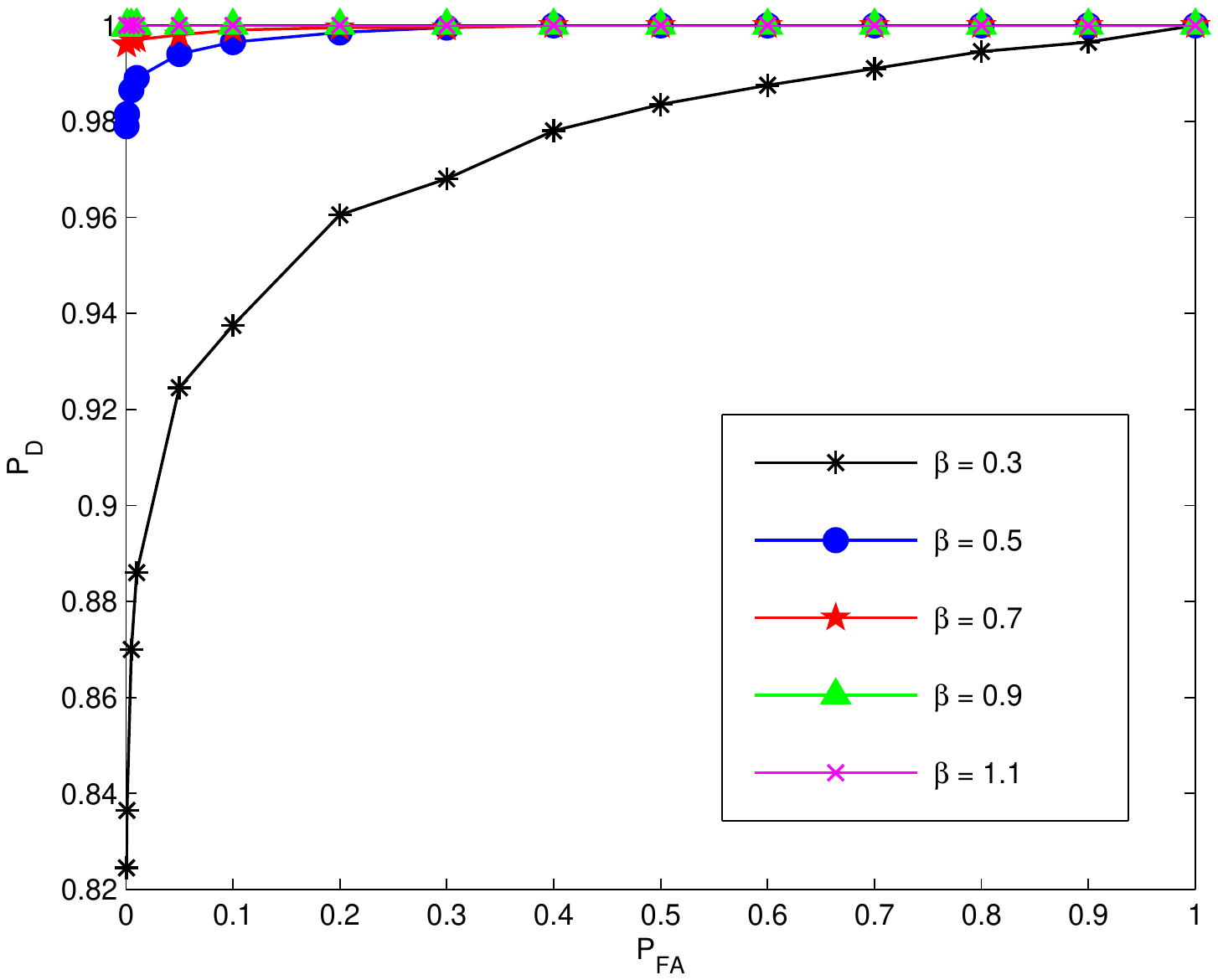}
\caption{ROC curves of periodogram-based sensing method under different $\beta$s.}
\label{fig:beta_diff}
\end{figure}

Figure \ref{fig:beta_diff} displays the ROC curves for $\mathrm{SNR} = -17$dB, $L = 11$, under different values of $\beta$. It is seen that the detector performance improves with $\beta$, which is consistent with the features of the periodogram of wireless microphone. Furthermore, for the values of $\beta$ beyond 0.5, the detection rate exceeds $98\%$ with the false alarm rage being as low as $1\%$.

\section{Augmented SCF-based Solution}
\label{sec:scf}

\subsection{Preliminary of SCF}

The SCF was originally proposed to reveal the cyclostationarity embedded in a signal, when its spectral components are temporally correlated \cite{gardner86:assp} \cite{gardner86:sp} \cite{gardner94:book}. The SCF of a continuous-time signal $x(t)$ is defined as \cite{gardner86:assp}
\begin{eqnarray}
\label{equ:syms_scf}
S^{\alpha}_x(f) &\triangleq& \lim_{T \rightarrow \infty} \lim_{\Delta t \rightarrow \infty} S^{\alpha}_{X_T}(f)_{\Delta t},\\
S^{\alpha}_{X_T}(f)_{\Delta t}  &\triangleq& \frac{1}{T \Delta t} \int^{\frac{\Delta t}{2}}_{-\frac{\Delta t}{2}} X_{T}\left(t, f + \frac{\alpha}{2}\right) X^\dag_{T}\left(t, f - \frac{\alpha}{2}\right) dt, \nonumber\\
X_T(t,f) &\triangleq& \int^{t + T/2}_{t - T/2} x(u)e^{-j2\pi fu}du, \nonumber
\end{eqnarray}
where $\alpha$ denotes the cycle frequency. Intuitively, the SCF (\ref{equ:syms_scf}) measures the normalized correlation between two spectral components of $x(t)$ at frequencies $f+ \alpha / 2$ and $f-\alpha/2$ over a length-$\Delta t$ time interval.

Given a sequence of discrete-time samples of signal $x[n], n = 1, 2, \ldots, N - 1$, a simple approximation of the SCF $S^\alpha_x(f)$ may be computed as \cite{gardner86:assp}\cite{sullivan95:sp}
\begin{equation}
\label{equ:sysm_discrete_scf}
\hat{S}^\alpha_x(f) \triangleq \frac{1}{MN} \sum^{(M-1)/2}_{k=-(M-1)/2}X(f+\frac{k}{N}+\frac{\alpha}{2})\cdot X^\dag(f+\frac{k}{N}- \frac{\alpha}{2}),
\end{equation}
where $M \leq N$ is an odd positive integer determining the frequency resolution $\Delta f = M/N$ of the estimate, and $X(f)$ is the discrete-time Fourier transform (DTFT) of $x[\cdot]$,
\begin{equation}
\label{equ:sysm_discrete_DFT}
X(f) \triangleq \sum^{N-1}_{n=0} x[n] e^{-j2\pi fn}.
\end{equation}
For efficiency, the computation of $\hat{S}^\alpha_x(f)$ is usually implemented by a fast Fourier transform (FFT) algorithm, over all of the values of $f$ such that $f \pm \alpha /2$ are integer multiples of $1/N$.

Most digitally modulated signals exhibit a certain degree of cyclostationarity, and have distinctive patterns in their SCFs. Stationary Gaussian noise, however, shows no cyclostationarity, so that for any $\alpha \neq 0$ its SCF is nearly zero. Therefore, methods based on SCF have been widely used for modulated signal detection and classification \cite{gardner86:sp}.

\subsection{Conjugate SCF and Augmented SCF}

When we apply SCF to CW and wireless microphone signals, a potential problem arises; that is, the SCF for $\alpha \neq 0$ is usually too weak to render any extractable feature for detection. This phenomenon may be illustrated by examining the SCF of a sampled bandpass CW: $x[n] = A \cos(2\pi f_c T_s n + \varphi)$. We denote $f_c T_s$ by $f_0$ as the normalized carrier frequency in discrete-time. For such a signal, its DTFT (\ref{equ:sysm_discrete_DFT}) computed using FFT at discrete frequencies $\{k/N\}_{k = 0}^{N - 1}$ is
\begin{eqnarray*}
&&\quad X(k/N) = \sum_{n = 0}^{N - 1} x[n] e^{-j 2\pi n k/N}\nonumber\\
&=& \frac{A}{2} e^{j\varphi} \frac{1 - e^{j2\pi f_0 N}}{1 - e^{2\pi ( f_0 - \frac{k}{N})}}
          + \frac{A}{2} e^{-j\varphi} \frac{1 - e^{-j2\pi f_0 N}}{1-e^{-2\pi ( f_0 + \frac{k}{N})}}
\end{eqnarray*}
from which we notice that a sharp spectral peak would occur only if for some $k$, $f_0 = \pm k/N$. Furthermore, from the approximation of the SCF in (\ref{equ:sysm_discrete_scf}), there are at most four locations on the $(\alpha, f)$-plane that exhibit sharp SCF feature peaks:
\begin{enumerate}
\item $\alpha = 0$, values of $f$ in the vicinity of $f \in \left[f_0 - \frac{M - 1}{2N}, f_0 + \frac{M - 1}{2N}\right]$ satisfying $f + k/N \pm \alpha/2 = f_0$ for some integer $k$;
\item $\alpha = 0$, values of $f$ in the vicinity of $f \in \left[- f_0 - \frac{M - 1}{2N}, - f_0 + \frac{M - 1}{2N}\right]$ satisfying $f + k/N \pm \alpha/2 = - f_0$ for some integer $k$;
\item $\alpha = 2 f_0$, values of $f$ in the vicinity of $f \in \left[- \frac{M - 1}{2N}, \frac{M - 1}{2N}\right]$ satisfying $f + k/N \pm \alpha/2 = \pm f_0$ for some integer $k$;
\item $\alpha = - 2 f_0$, values of $f$ in the vicinity of $f \in \left[- \frac{M - 1}{2N}, \frac{M - 1}{2N}\right]$ satisfying $f + k/N \pm \alpha/2 = \mp f_0$ for some integer $k$.
\end{enumerate}

Now in order to exhibit more features on the $(\alpha, f)$-plane for the purpose of detection, we propose to introduce the concepts of conjugate SCF and augmented SCF. The idea of conjugate SCF is simply swapping the role of $f$ and $\alpha$, to define
\begin{equation}
\label{equ:sysm_discrete_scf_conj}
\hat{S}^\alpha_x(f)_c \triangleq \frac{1}{MN} \sum^{(M-1)/2}_{k=-(M-1)/2}X(f+\frac{k}{N}+\frac{\alpha}{2})\cdot X^\dag(f-\frac{k}{N}- \frac{\alpha}{2})
\end{equation}
where the subscript $c$ represents ``conjugate''. The reader may compare its difference from the definition of the SCF (\ref{equ:sysm_discrete_scf}). Compared with SCF, The conjugate SCF appears to exhibit more features along the $\alpha$ axis. For the sampled bandpass CW example examined above, the following four locations on the $(\alpha, f)$-plane possibly exhibit sharp conjugate SCF feature peaks:
\begin{enumerate}
\item $f = 0$, values of $\alpha$ in the vicinity of $\alpha \in \left[2f_0 - \frac{M - 1}{N}, 2f_0 + \frac{M - 1}{N}\right]$ satisfying $f \pm (k/N + \alpha/2) = \pm f_0$ for some integer $k$;
\item $f = 0$, values of $\alpha$ in the vicinity of $\alpha \in \left[- 2f_0 - \frac{M - 1}{N}, - 2f_0 + \frac{M - 1}{N}\right]$ satisfying $f \pm (k/N + \alpha/2) = \mp f_0$ for some integer $k$;
\item $f = f_0$, values of $\alpha$ in the vicinity of $\alpha \in \left[- \frac{M - 1}{N}, \frac{M - 1}{N}\right]$ satisfying $f \pm (k/N + \alpha/2) = f_0$ for some integer $k$;
\item $f = - f_0$, values of $\alpha$ in the vicinity of $\alpha \in \left[- \frac{M - 1}{N}, \frac{M - 1}{N}\right]$ satisfying $f \pm (k/N + \alpha/2) = - f_0$ for some integer $k$.
\end{enumerate}

\begin{figure}
\center
\includegraphics[width=3in]{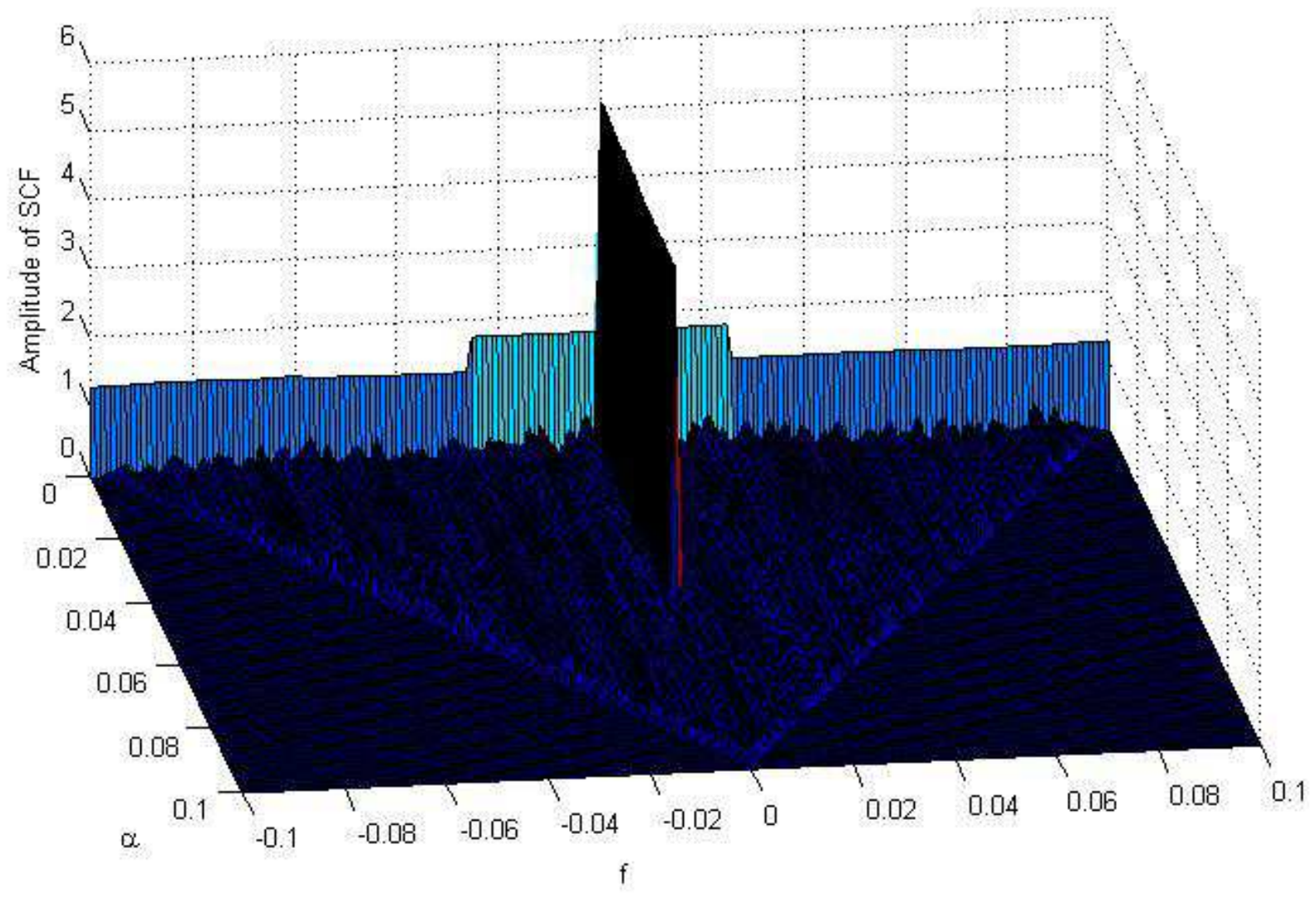}
\caption{Augmented SCF of a CW in baseband. The frequency $f$ and the cycle frequency $\alpha$ are normalized (the same in subsequent figures of augmented SCFs).}\label{fig:sysm4}
\end{figure}

\begin{figure}
\center
\includegraphics[width=3in]{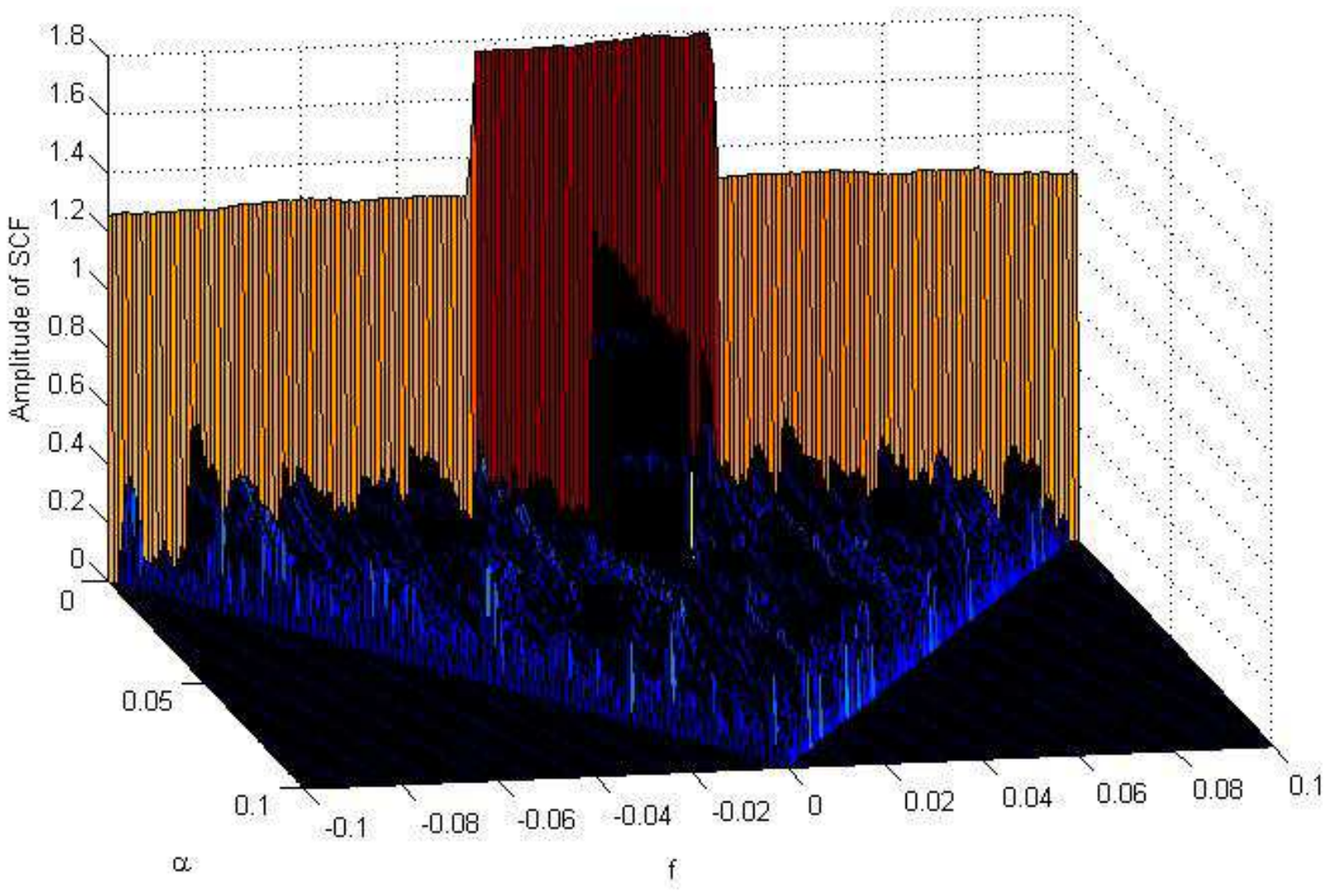}
\caption{Augmented SCF of a wireless microphone signal in baseband.}\label{fig:sysm5}
\end{figure}

Now we further define the augmented SCF as
\begin{eqnarray}
\hat{S}^\alpha_x(f)_a \triangleq \kappa_1 \left.\hat{S}^{\alpha}_x(f)\right|_{\alpha = 0} + \kappa_2 \left.\hat{S}^{\alpha}_x(f)_c\right|_{\alpha \neq 0},
\end{eqnarray}
where the subscript $a$ represents ``augmented''. The weighting factors $\kappa_{1, 2}$ balance the impacts from SCF and the conjugate SCF. Note that, for SCF we only keep its components on $\alpha = 0$, so indeed $\hat{S}^{\alpha = 0}_x(f)$ is an estimate of the signal PSD; while for the conjugate SCF we ignore its components for $\alpha = 0$. Such a definition of the augmented SCF here is specifically tailored to suit the problem of distinguishing between CW and wireless microphone signals, while other definitions are possible (and perhaps more appropriate) for more extensive potential applications. Figures \ref{fig:sysm4} and \ref{fig:sysm5} display the magnitudes of the augmented SCFs of a baseband CW and a baseband wireless microphone signal, respectively. In both figures we use the same set of weighting factors $\kappa_{1, 2}$. Visually we can notice the difference between the two augmented SCFs, in that the CW tends to have larger conjugate SCF magnitudes compared with the SCF at $\alpha = 0$, while the wireless microphone signal shows an opposite behavior. A precise mathematical explanation of such a difference does not appear straightforwardly available, given our limited knowledge regarding the statistical properties of wireless microphone signals. In Appendix \ref{app:aug-scf} we sketch out a first-cut analysis which provides useful intuition about the underlying mechanism.

\subsection{Method Description}

Having noticed the visual difference between the augmented SCFs of a CW and a wireless microphone signal, we need to quantitatively convey such difference using a decision statistic. In this work, we choose to use the following statistic,
\begin{eqnarray}
T_a \triangleq 1 - \frac{|\Psi|\sum_{\alpha\in \Omega} \left.|\hat{S}^\alpha_x (f)_a|\right|_{f = 0}} {|\Omega|\sum_{f\in\Psi} \left.|\hat{S}^{\alpha}_x (f)_a|\right|_{\alpha = 0}}.
\end{eqnarray}
Herein, $\Psi$ is a window of frequency bins among which we average $|\hat{S}^{\alpha=0}_x (f)_a|$, and $\Omega$ is a window of cycle frequency bins among which we average $|\hat{S}^\alpha_x (f=0)_a|$. So the statistic $T_a$ is just one minus the ratio between the averaged augmented SCF magnitudes on $f = 0$ and the averaged augmented SCF magnitudes on $\alpha = 0$. From our visual observation, for CW, $T_a$ tends to be small, while for wireless microphone signals, $T_a$ tends to be large.

We remark that, in order to compute $T_a$, we only need to compute the values of the SCF on $\alpha = 0$ (that is, the PSD) within window $\Psi$, and the values of the conjugate SCF on $f = 0$ within window $\Omega$. Without the need to compute the augmented SCF over the entire $(\alpha, f)$-plane, the implementation complexity of the detection method may be moderate and suitable for practice.

Even under $\mathcal{H}_0$, the distribution of $T_a$ is complicated to analyze. Hence we use Monte Carlo simulation to determine the decision threshold $\gamma$, in order to meet a specified target false alarm rate. The decision rule is as follows,
\begin{eqnarray*}
\hat{\mathcal{H}} = \mathcal{H}_0 \quad\mbox{if}\; T_a \leq \gamma;\; \mbox{and}\; \mathcal{H}_1 \quad\mbox{otherwise}.
\end{eqnarray*}
In order to combat against noise and thus improve the decision accuracy, the statistics of multiple separate segments of signals may be averaged or appropriately combined.

\subsection{Simulation Results}

The simulation setup is identical to that in Section \ref{subsec:psd-simu}, except that for computing the augmented SCF the sampling rate is set as $20$MHz. The weighting factors in $\hat{S}^\alpha_x(f)_a$ are chosen as $\kappa_1 = 0.1$ and $\kappa_2 = 1$. The $T_a$ statistics of five separate segments of signals each of time duration $5$ms are averaged to improve the decision accuracy. The window sizes of $\Psi$ and $\Omega$ are chosen as five and ten, respectively.

\begin{figure}
\center
\includegraphics[width=3.5in]{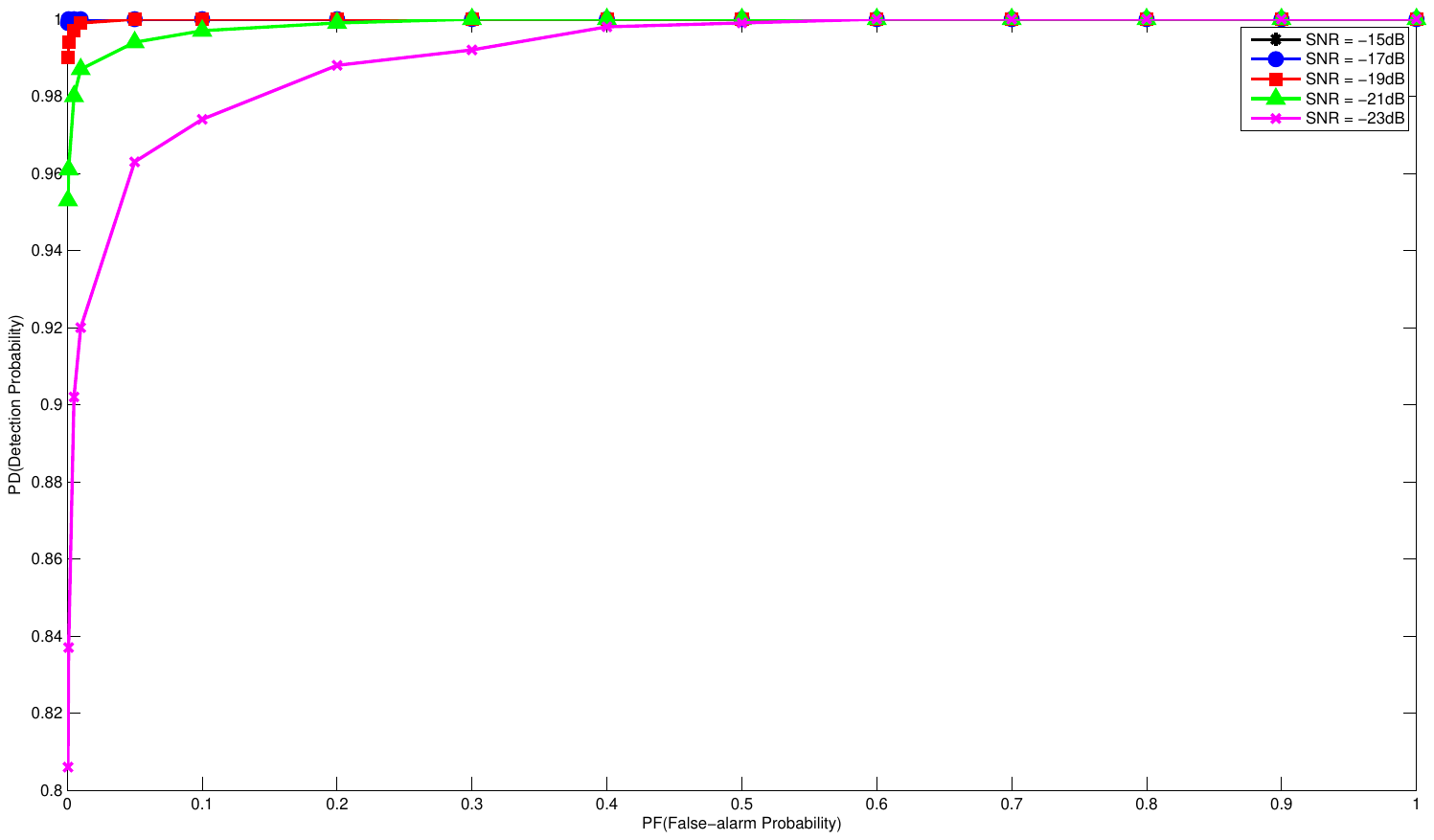}
\caption{ROC curves of augmented SCF-based sensing method under different SNRs.}\label{fig:sysm6}
\end{figure}

Figure \ref{fig:sysm6} displays the ROC curves of the detection method, for $\beta = 2$, under different values of $\mathrm{SNR}$. It is clearly seen that increasing the SNR dramatically improves the detector performance. When the SNR is (or higher than) $-21$dB, the detection rate exceeds 98\% with the false alarm rate being $5\%$.

\begin{figure}
\center
\includegraphics[width=3.5in]{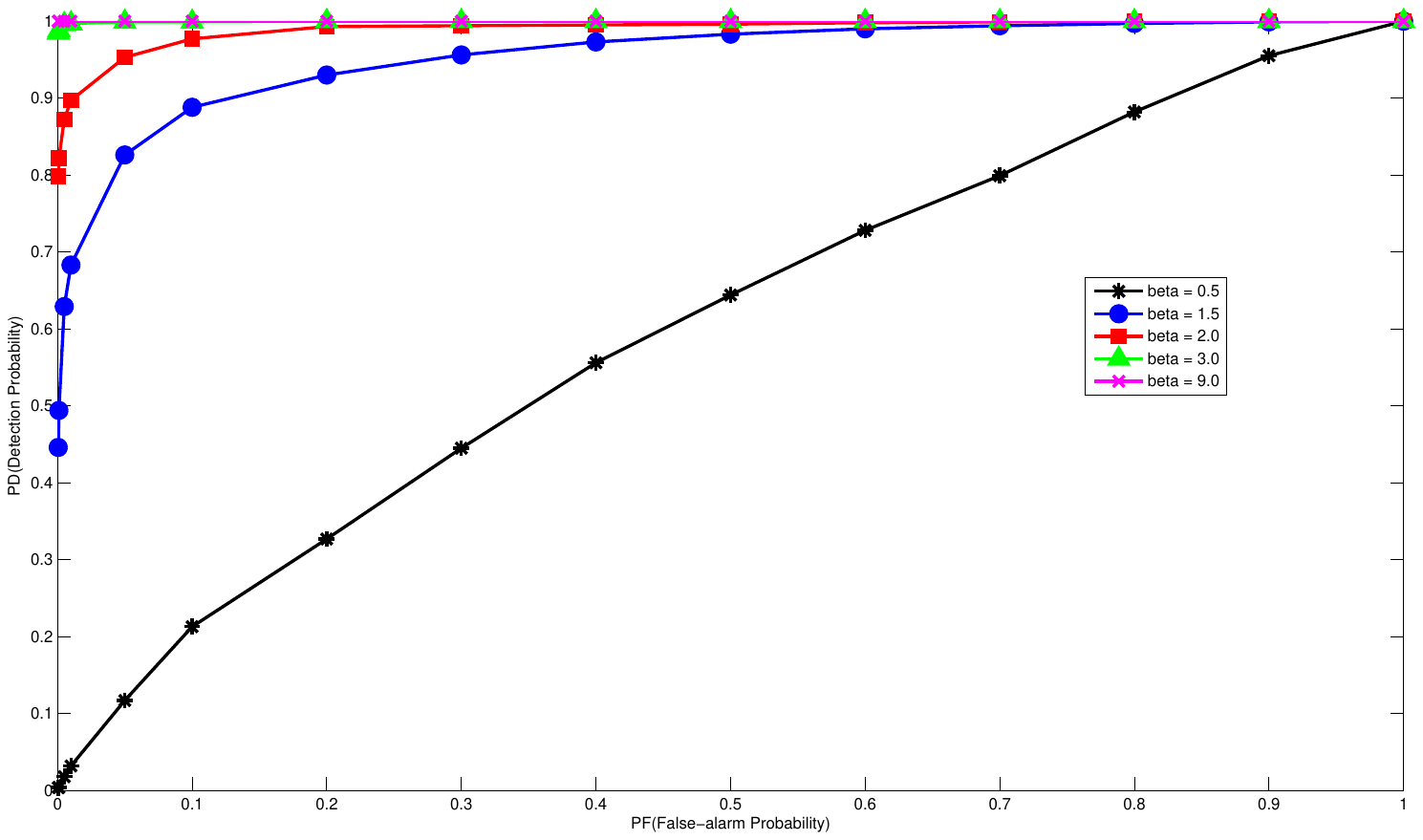}
\caption{ROC curves of augmented SCF-based sensing method under different $\beta$s.}\label{fig:sysm7}
\end{figure}

Figure \ref{fig:sysm7} displays the ROC curves for $\mathrm{SNR} = -23$dB, under different values of $\beta$. Even for such an extremely low SNR, the detection remains essentially error-free over a wide range of $\beta$s until $\beta$ gets as small as $2$. On the other hand, in our simulation we observe a somewhat unexpected phenomenon that, if $\beta$ grows too large, say, several tens (not displayed in the figures herein), then the bandwidth dispersion in the wireless microphone signal becomes so severe that even the center peak in the PSD becomes diminished, making the denominator in $T_a$ small so as to decrease the value of $T_a$, and thus the detection rate drops.

\section{Experimental Validation}
\label{sec:experiment}

We present in this section experimental validation of the proposed sensing methods.

\subsection{Experiment System Description}

\begin{figure}[!t]
\centering
\includegraphics[width=4in]{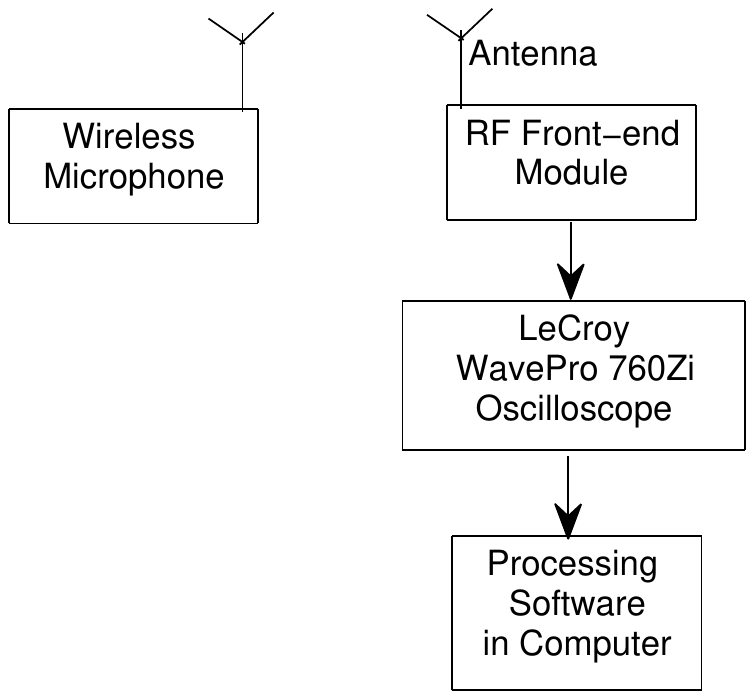}
\caption{Schematic structure of sensing experiment setup.}
\label{fig:system_module}
\end{figure}

\begin{figure}[!t]
\centering
\includegraphics[width=3in]{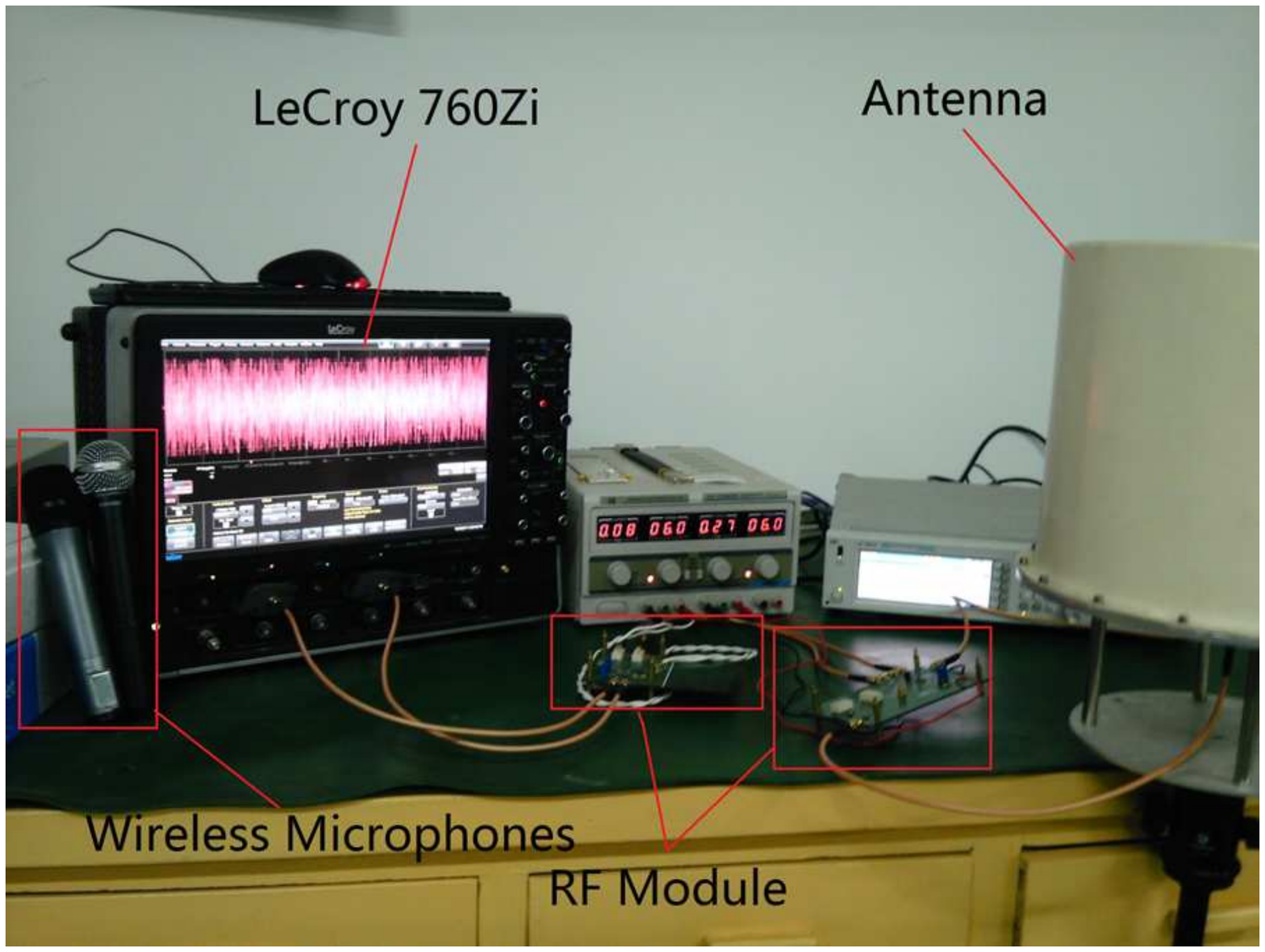}
\caption{Experiment test bench of wireless microphone sensing.}
\label{fig:exp_module}
\end{figure}

Figure \ref{fig:system_module} schematically illustrates the structure the experiment setup, and Figure \ref{fig:exp_module} shows the experiment test bench. The experiment system consists of a wireless microphone, an antenna, a RF front-end module, and an oscilloscope. In experiments we use two makes of wireless microphones: one is Sennheiser EW100G2 which is capable of transmitting on $1440$ frequency points (with a stepsize of $25$kHz) between $786$MHz and $822$MHz, and the other is Shure UR8D which is capable of transmitting on $200$ frequency points (with a stepsize of $160$kHz) between $786$MHz and $822$MHz.\footnote{In China it has not been allowed to transmit on TV bands, so we choose to transmit on the spectrum just above the TV band upper edge, where the interference characteristics are quite similar as in TV bands.} The signals from a wireless microphone are received by the antenna, passed through the RF front-end module, and then fed to the oscilloscope where they are recorded and stored, before being processed by the computer program which implements the proposed sensing methods as described in the previous sections.

\subsection{Experiment Results}

\begin{figure}[!t]
\centering
\includegraphics[width=2.8in]{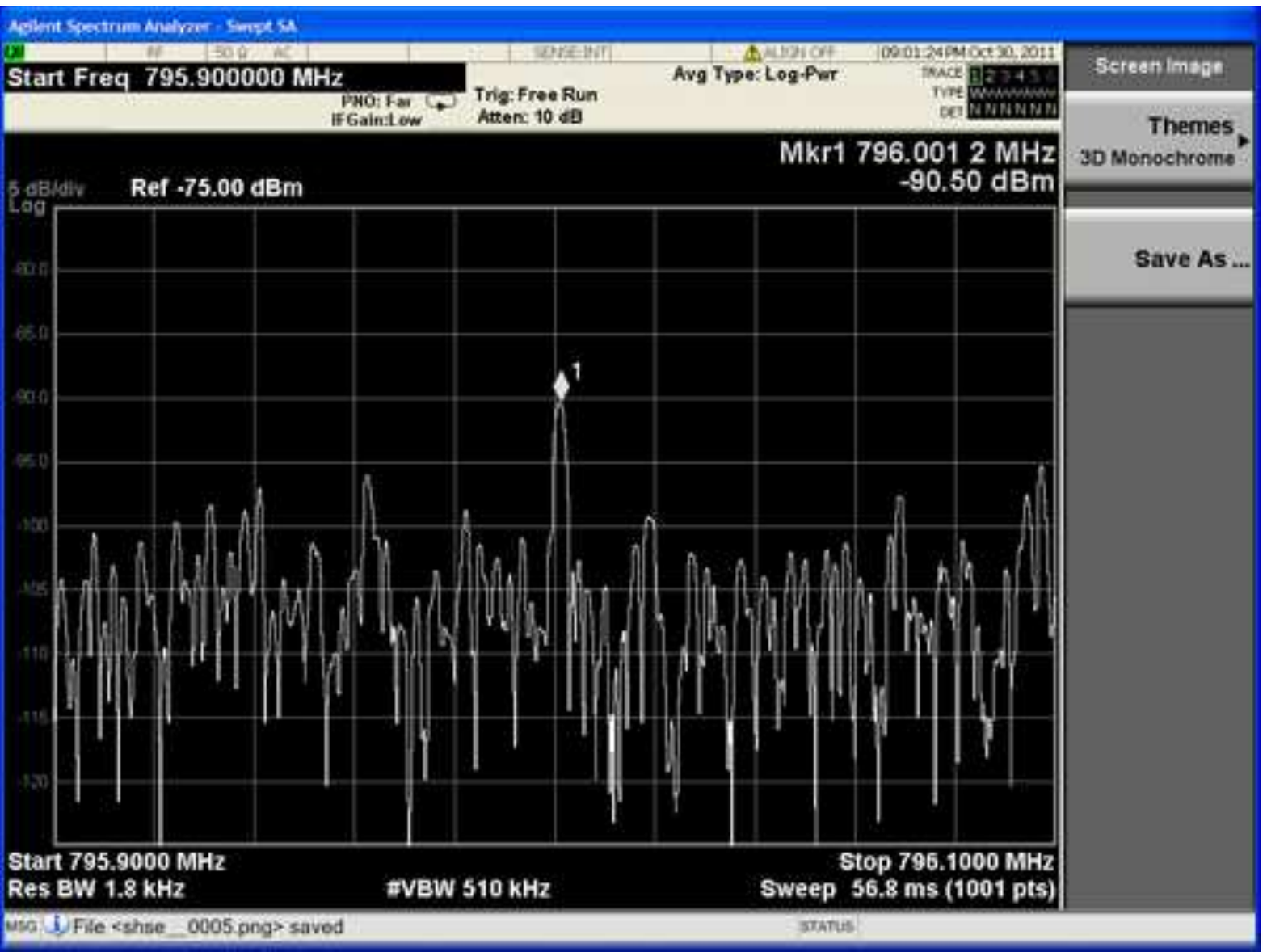}
\caption{Measured spectrum of Sennheiser EW100G2.}
\label{fig:shse}
\end{figure}

\begin{figure}[!t]
\centering
\includegraphics[width=2.8in]{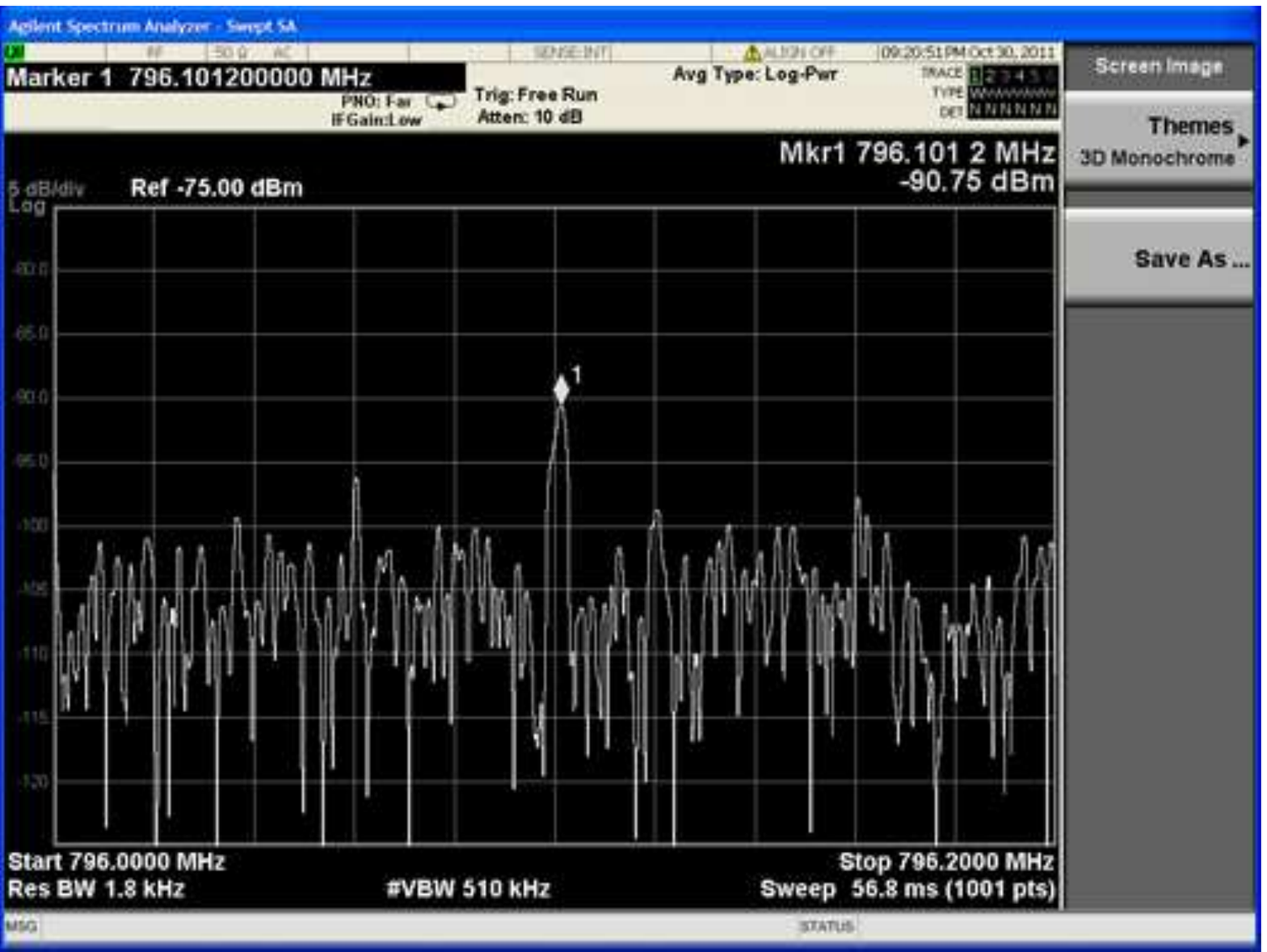}
\caption{Measured spectrum of Shure UR8D.}
\label{fig:shure}
\end{figure}

We collected signal data using the wireless microphones indoors in a multi-floor building (Department of Electronic Engineering and Information Science in the west campus of University of Science and Technology of China), at a distance so as to sufficiently attenuate the received signal power. The wireless microphones were tested in both silent and voiced modes. In the silent mode there was no audio input to the wireless microphones, and in the voiced mode there was a recorded music playback to feed the wireless microphones at moderate volume. For each make of wireless microphone in each mode, twenty experiments were conducted. Figures \ref{fig:shse} and \ref{fig:shure} are examples of the captured spectra of Sennheiser EW100G2 and Shure UR8D in silent mode, respectively, on a spectrum analyzer. Because the background noise power density in the spectrum analyzer was measured to be about $-139$dBm/Hz (as indicated in the spectrum analyzer datasheet, and also can be confirmed from the figures taking into account the resolution bandwidth of $1.8$kHz), substantially higher than the ideal thermal noise density $-174$dBm/Hz, the wireless microphone signal power was set to roughly $-90$dBm, translating into a SNR approximately $-20$dB over $8$MHz TV channel bandwidth. Similar exercise was also conducted when recording wireless microphone signals in the oscilloscope.

In the experiments, we also tested the false alarm performance, by running the sensing methods for a number of noticeable narrowband interferences in the received spectrum. Every time such a narrowband interference being decided as a wireless microphone signal, we interpreted it as a false alarm event.

\subsubsection{Results of Periodogram-based Sensing Method} For Sennheiser EW100G2, in silent mode the detection rate $P_\mathrm{D}$ was $100\%$, against the false alarm rate of $P_\mathrm{FA} \approx 8\%$; while in voiced mode, $P_\mathrm{D}$ was $100\%$ against $P_\mathrm{FA} \approx 5\%$. For Shure UR8D, in silent mode $P_\mathrm{D}$ was $100\%$ against $P_\mathrm{FA} \approx 4\%$; while in voiced mode, $P_\mathrm{D}$ was $100\%$ against $P_\mathrm{FA} < 1\%$. The trend in these results is consistent with the simulation in Section \ref{subsec:psd-simu}.

\subsubsection{Results of Augmented SCF-based Sensing Method} 
\begin{figure}
\centering
\subfigure[Sennheiser EW100G2 (silent)]{
\includegraphics[width=1.5in]{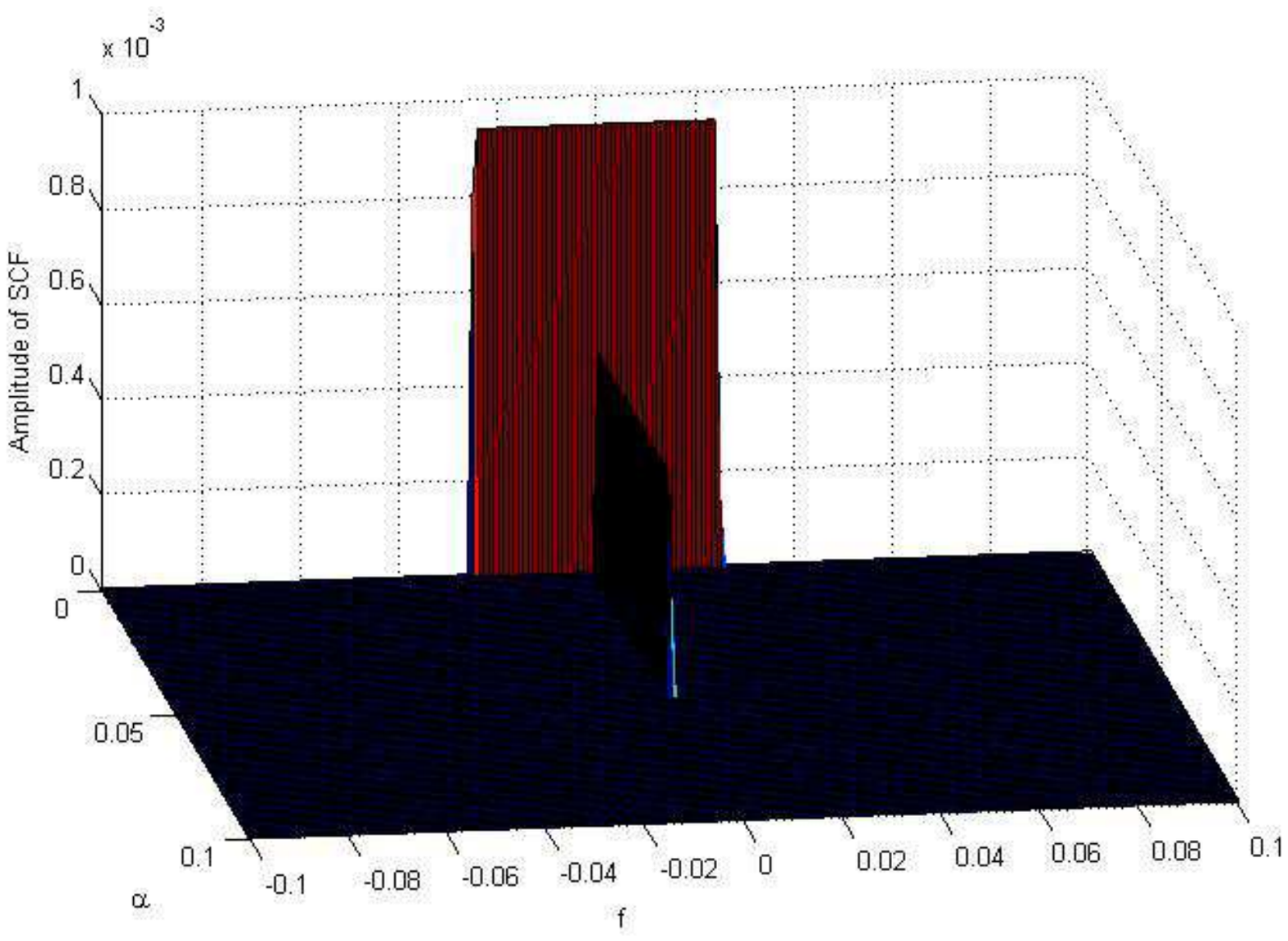} }
\subfigure[Sennheiser EW100G2 (voiced)]{
\includegraphics[width=1.5in]{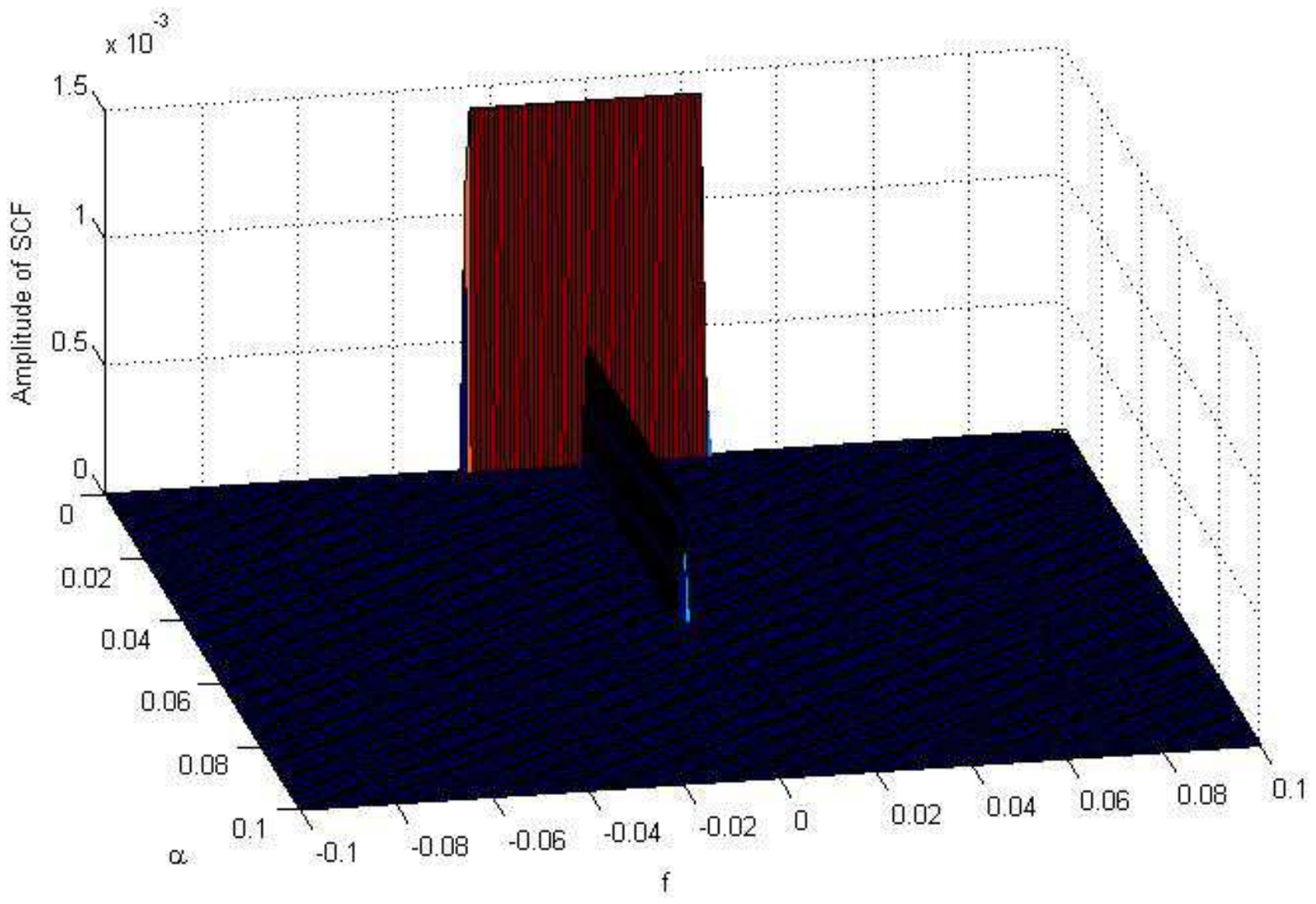}}
\subfigure[Shure UR8D (silent)]{
\includegraphics[width=1.5in]{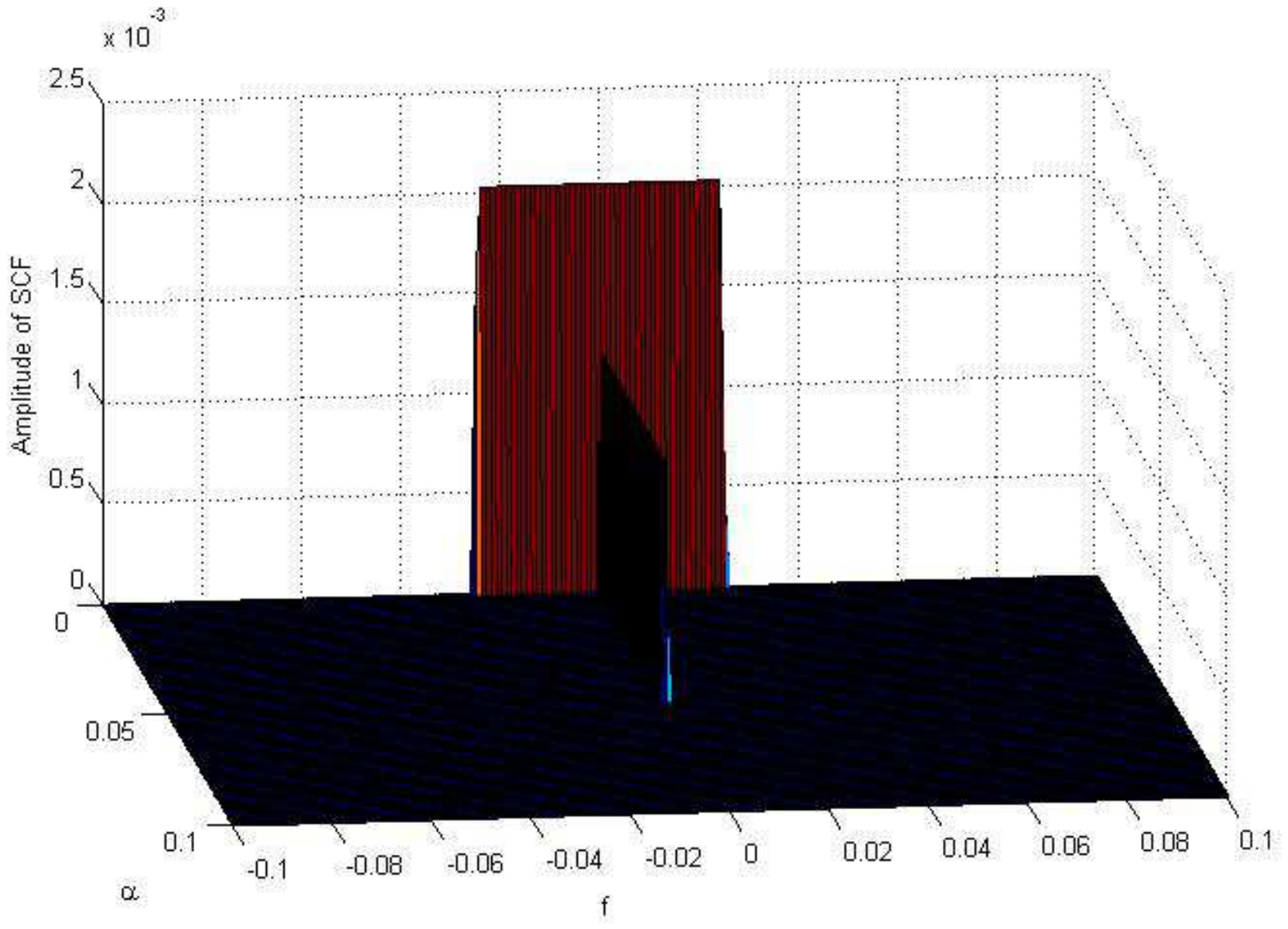} }
\subfigure[Shure UR8D (voiced)]{
\includegraphics[width=1.5in]{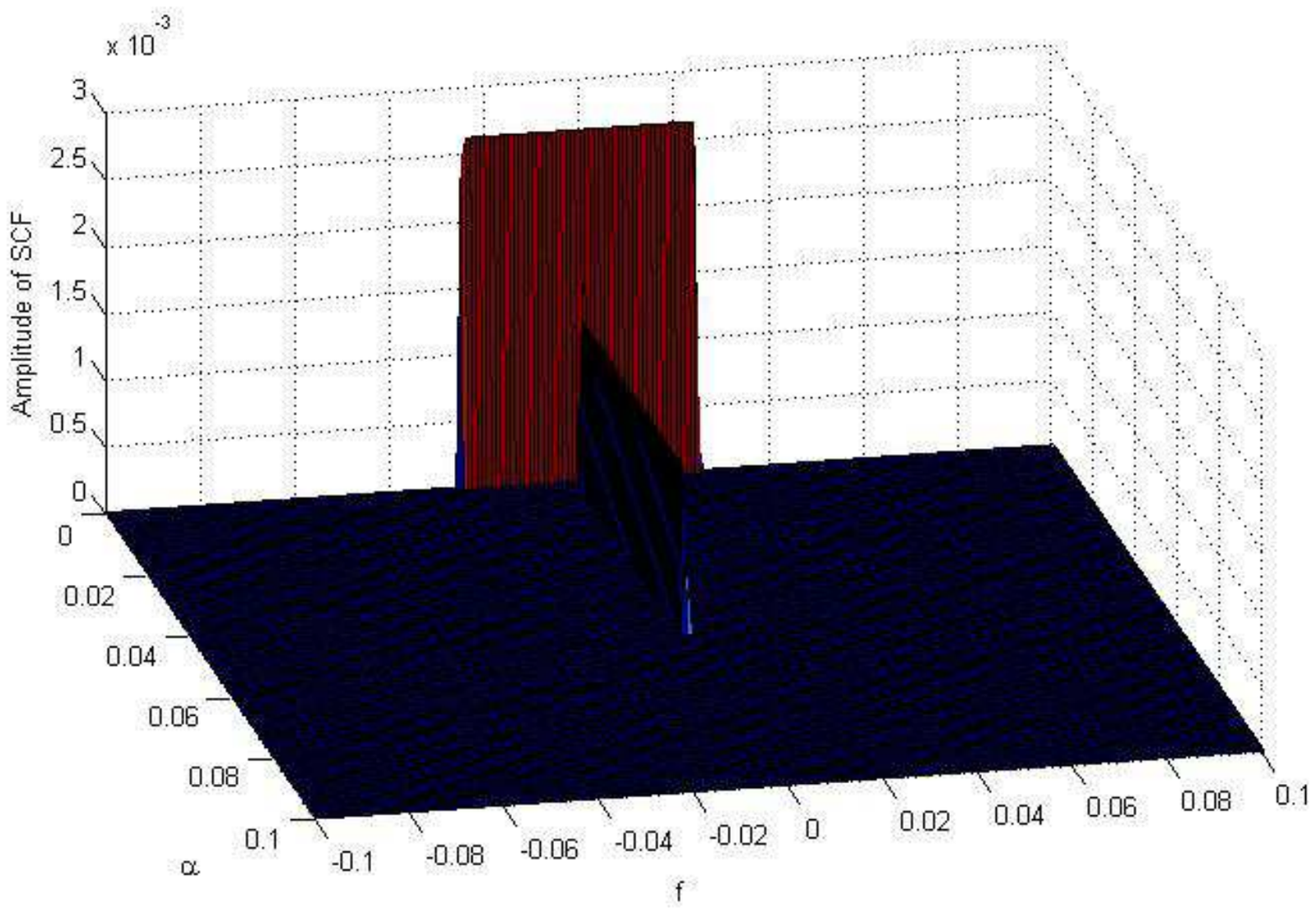}}
\subfigure[CW-like narrowband interference]{
\includegraphics[width=1.5in]{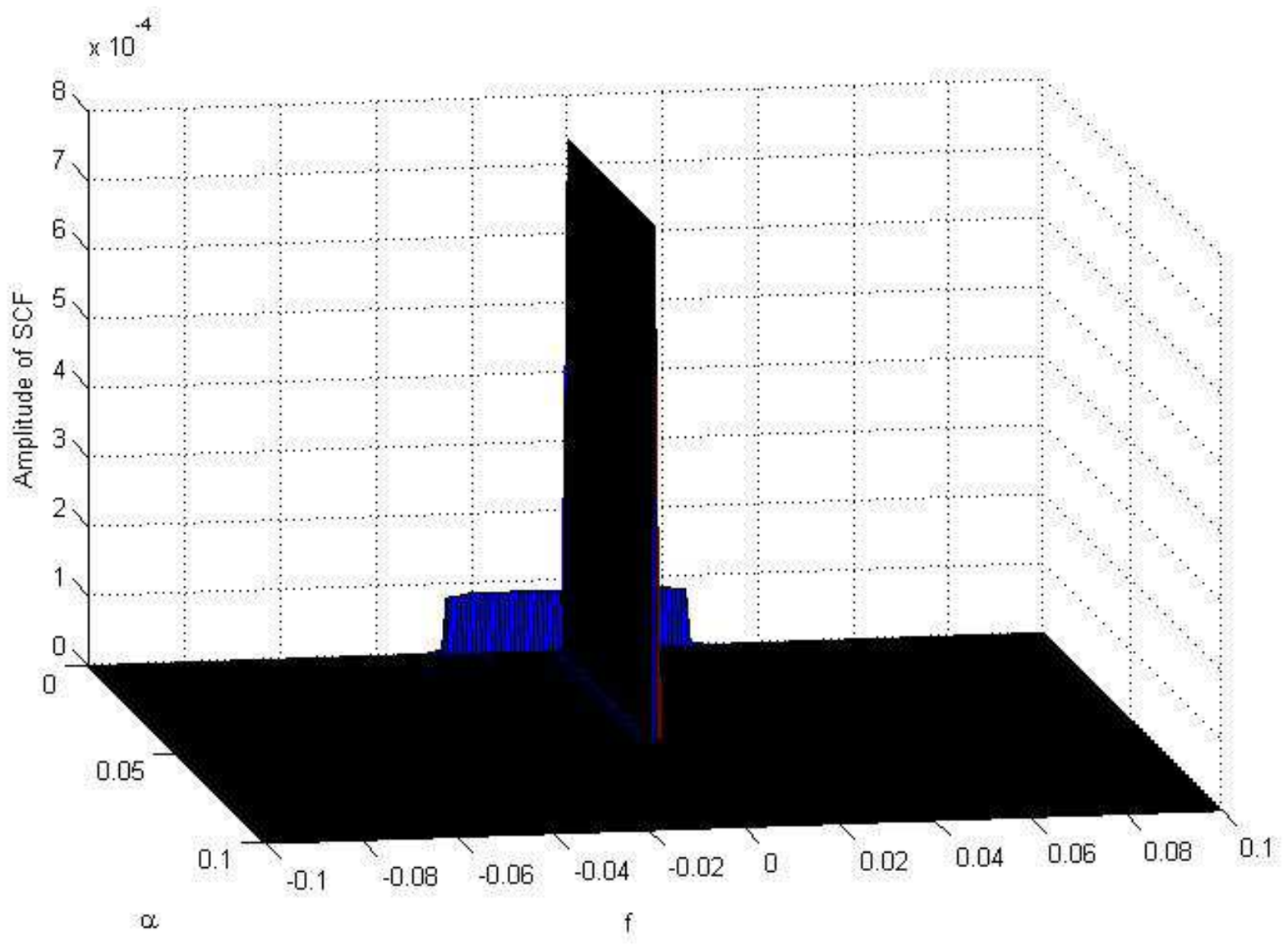}}
\caption{Augmented SCFs of different makes of wireless microphones and CW-like interference.}
\label{fig:exp_ssss}
\end{figure}

For each make of wireless microphone in each mode, $P_\mathrm{D}$ was $100\%$ without an occurrence of a false alarm event, throughout the twenty experiments. The augmented SCFs under different situations are displayed in Figure \ref{fig:exp_ssss} for visual illustration. We clearly observe that, the shape of augmented SCF effectively captures the key difference between FM wireless microphones and CW-like narrowband interferences.

\section{Conclusion}
\label{sec:conclusion}

We considered wireless microphone sensing in TVWS, treating its core problem of distinguishing between CW-like narrowband interference and FM wireless microphone signals. We developed two solutions, one based on periodogram and the other based on a proposed augmented SCF, to deal with this problem. Both the two solutions are logical consequences of our perceptual observations of the key features underlying CW and wireless microphone signals. Both simulation results and experimental validation results indicate that the proposed solutions are promising for reliably sensing wireless microphones in TVWS.

There are a number of items in future work. First, for the proposed solutions, further analyzing their theoretical properties and computational complexities is important towards a full understanding of their potential and limitation. Second, for applying those solutions to real systems, hardware-oriented implementation techniques need to be studied and developed, and the developed experimental system needs to experience more extensive field tests, in order to fully validate the concepts.

\section*{Acknowledgement}
The research was supported in part by the National Natural Science Foundation of China (61071095), the Fundamental Research Funds for the Central Universities, and 100 Talents Program of Chinese Academy of Sciences.

\newpage
\section*{Appendix}

\subsection{Statistical Properties of CW Periodogram and Frequency Offset Correction}
\label{app:off-est}

With a frequency offset $\Delta f = f_c - \hat{f}_c$, the down-converted baseband signal of CW can be written as
\begin{eqnarray}
x[n] = \frac{A}{2} e^{-j\varphi} e^{j2\pi n T_s \Delta f} + w[n].
\end{eqnarray}
For a length-$N$ segment of $x[\cdot]$, after taking discrete Fourier transform (DFT), we obtain
\begin{eqnarray}
X[k] &=& \frac{1}{\sqrt{N}} \sum_{n = 0}^{N - 1} x[n] e^{-j2\pi kn/N}\nonumber\\
&=& \frac{A}{2\sqrt{N}} \sum_{n = 0}^{N - 1} e^{-j\varphi + j2\pi n \left(\Delta f T_s - k/N\right)} +\nonumber\\
&& \quad\quad \frac{1}{\sqrt{N}} \sum_{n = 0}^{N - 1} w[n] e^{-j2\pi kn/N}\nonumber\\
&=& \frac{A}{2\sqrt{N}} e^{-j\varphi} e^{j\pi (N-1)\zeta[k]} \mathrm{sad}(\pi\zeta[k], N) +\nonumber\\
&& \quad\quad \frac{1}{\sqrt{N}} \sum_{n = 0}^{N-1} w[n] e^{-j2\pi kn/N}\nonumber\\
&\triangleq& X^\prime[k] + W[k],\quad\mbox{for}\; k = 0, \cdots, N-1,
\end{eqnarray}
where $\mathrm{sad}(\pi\zeta[k], N) \triangleq \frac{\sin(\pi N\zeta[k])}{\sin(\pi\zeta[k])}$ and $\zeta[k] \triangleq T_s \Delta f - k/N$. We also note that $\mathrm{sad}(0, N) = \lim_{\zeta \rightarrow 0} \mathrm{sad}(\pi\zeta, N) = N$. The periodogram of $x[n]$ is
\begin{eqnarray}
&& \quad \xi[k] \triangleq |X[k]|^2 = \nonumber\\
&& |X^\prime[k]|^2 + X^\prime[k] W^\dag[k] + X^{\prime\dag}[k] W[k] + |W[k]|^2.
\end{eqnarray}
Due to the modeling assumption on noise, we have $W[k] \sim \mathcal{CN}(0, \sigma^2)$, and consequently the mean and variance of $\xi[k]$ are respectively
\begin{eqnarray}
\label{mean_X}
\mathbb{E}\left\{ \xi[k] \right\} &=& \sigma^2 \left[\frac{\mathrm{SNR}}{N} \mathrm{sad}^2(\pi\zeta[k], N) + 1\right]\\
\label{var_X}
\mathrm{var}\left\{ \xi[k] \right\} &=& \sigma^4 \left[\frac{2 \mathrm{SNR}}{N} \mathrm{sad}^2(\pi\zeta[k], N) + 1\right].
\end{eqnarray}

In order to suppress the effect of noise, we usually smooth out $\xi[\cdot]$ by averaging $M$ segments of such length-$N$ periodograms to obtain an improve estimate of the PSD of the received signal, as $\bm{\xi}_a = (\xi_a[0],\; \xi_a[1],\; \ldots, \xi_a[N - 1])$ where the subscript $a$ represents averaging over $M$ segments.

Below is the procedure we used in our work to estimate the frequency offset $\Delta f$. First, we identify the largest two elements in $\bm{\xi}_a$ as $\xi_a[k_1]$ and $\xi_a[k_2]$, $k_1 < k_2$, respectively. Then, by treating $\xi_a[k]$ as the mean $\mathbb{E}\left\{\xi[k]\right\}$ in (\ref{mean_X}), we get
\begin{eqnarray}
\frac{\xi_a[k_1]}{\sigma^2} - 1 &=& \frac{\mathrm{SNR}}{N} \mathrm{sad}^2(\pi\zeta[k_1], N)\\
\frac{\xi_a[k_2]}{\sigma^2} - 1 &=& \frac{\mathrm{SNR}}{N} \mathrm{sad}^2(\pi\zeta[k_2], N).
\end{eqnarray}
Taking their ratio leads to
\begin{eqnarray}
\label{eqn:deltafeqn}
\left[\frac{\mathrm{sad}(\pi\zeta[k_1], N)}{\mathrm{sad}(\pi\zeta[k_2], N)}\right]^2 = \frac{\xi_a[k_1] - \sigma^2}{\xi_a[k_2] - \sigma^2}.
\end{eqnarray}
Now solving the equation (\ref{eqn:deltafeqn}) yields an estimate of $\Delta f$. We then correct the down-conversion frequency to $\hat{f}_c + \Delta f$ which would be close to $f_c$, and perform the down-conversion procedure once again to come up with the hypothesis testing problem (\ref{eqn:dt-hypo-1})(\ref{eqn:dt-hypo-0}). Despite its simplicity, this correction procedure has shown satisfying performance in both simulation and experiments.

\subsection{The Detection Mechanism of Augmented SCF}
\label{app:aug-scf}

For simplicity, we ignore the noise and write the signal vectors in the frequency domain of CW and FM wireless microphone signals as
\begin{equation}
\label{equ:app_cw}
X[k] = \left\{ \begin{array}{ll}
Ae^{j\varphi}, & k = 0 \\
0, & k=1,\ldots,N-1, \mbox{and}
\end{array} \right.
\end{equation}
\begin{equation}
\label{equ:app_wm}
X[k] = \left\{ \begin{array}{ll}
A_k e^{j\varphi_k}, & k \in [0,L-1]\cup[N-L,N-1] \\
0, & k\in [L,N-L-1],
\end{array} \right.
\end{equation}
respectively. Here $L$ denotes one half the size of nonzero frequency bins in the frequency domain. We pick the parameter $M$ in the definitions of SCF and conjugate SCF such that $L < \frac{M-1}{2}$ holds.

For CW, the magnitudes of its augmented SCF can be explicitly evaluated as
\begin{equation}
\label{equ:app_cwscf}
|\hat{S}^{\alpha}_x(f)_a| = \left\{ \begin{array}{ll}
\kappa_1\frac{A^2}{M N}, & \alpha=0,f \in [-\frac{M-1}{2N},\frac{M-1}{2N}] \\
\kappa_2\frac{A^2}{M N}, & f=0,\alpha \in [-\frac{M-1}{N},\frac{M-1}{N}]\\
0, & \mathrm{others}.
\end{array} \right.
\end{equation}

For FM wireless microphone signal, the situation is different since there are more than one nonzero frequency bins in the frequency domain. The magnitudes of its augmented SCF can be written as shown in the following,
\begin{equation}
\label{equ:app_wmscf}
|\hat{S}^{\alpha}_x(f)_a| = \left\{ \begin{array}{l}
\frac{\kappa_1}{M N} \sum^{Nf+\frac{M-1}{2}}_{k=-L} {A^2_k},\\ \quad\quad \alpha=0, f \in [-\frac{L}{N}-\frac{M-1}{2N},\frac{L}{N}-\frac{M-1}{2N}] \\
\frac{\kappa_1}{M N} \sum^{L}_{k=-L} {A^2_k},\\ \quad\quad \alpha=0, f \in [\frac{L}{N}-\frac{M-1}{2N},-\frac{L}{N}+\frac{M-1}{2N}] \\
\frac{\kappa_1}{M N} \sum^{L}_{k=Nf-\frac{M-1}{2}} {A^2_k},\\ \quad\quad \alpha=0, f \in [-\frac{L}{N}+\frac{M-1}{2N},\frac{L}{N}+\frac{M-1}{2N}] \\
\frac{\kappa_2}{M N} |\sum^{\frac{N\alpha}{2}+\frac{M-1}{2}}_{k=-L} {A_{k} A_{-k} e^{j(\varphi_{k} - \varphi_{-k})}}|,\\ \quad\quad f=0, \alpha \in [-\frac{2L}{N}-\frac{M-1}{N},-\frac{2L}{N}+\frac{M-1}{N}] \\
\frac{\kappa_2}{M N} |\sum^{L}_{k=-L} {A_{k} A_{-k} e^{j(\varphi_{k} - \varphi_{-k})}}|,\\ \quad\quad f=0, \alpha \in [-\frac{2L}{N}+\frac{M-1}{N},\frac{2L}{N}-\frac{M-1}{N}]\\
\frac{\kappa_2}{M N} |\sum^{L}_{k=\frac{N\alpha}{2}-\frac{M-1}{2}} {A_{k} A_{-k} e^{j(\varphi_{k} - \varphi_{-k})}}|,\\ \quad\quad f=0, \alpha \in [\frac{2L}{N}-\frac{M-1}{N},\frac{2L}{N}+\frac{M-1}{N}] \\
0, \quad\mathrm{others}.
\end{array} \right.
\end{equation}

Now for simplicity let us set $\Psi=[\frac{L}{N}-\frac{M-1}{2N}, -\frac{L}{N}+\frac{M-1}{2N}]$, $\Omega = [-\frac{2L}{N}+\frac{M-1}{N}, \frac{2L}{N}-\frac{M-1}{N}]$, and $\kappa_1 = \kappa_2 = 1$. In view of (\ref{equ:app_cwscf}) for CW, we notice that the averaged magnitudes of its SCF (over $\alpha=0$) and its conjugate SCF (over $f = 0$) are identical. In view of (\ref{equ:app_wmscf}) for FM wireless microphone, however, its SCF (over $\alpha = 0$) and its conjugate SCF (over $f = 0$) show different behaviors. That is, the magnitudes of the SCF (over $\alpha = 0$) within the window $\Psi$ are identically $\frac{\kappa_1}{M N} \sum^{L}_{k=-L} {A^2_k}$, while the magnitudes of the conjugate SCF (over $f = 0$) within the window $\Omega$ are $\frac{\kappa_2}{M N} |\sum^{L}_{k=-L} {A_{k} A_{-k} e^{j(\varphi_{k} - \varphi_{-k})}}|$. Due to the random phase incoherences $e^{j(\varphi_{k} - \varphi_{-k})}$ therein, empirically the conjugate SCF is expected to be much smaller than the SCF. In order to further enhance such a difference, in our detection method we adjust the weighting factors $\kappa_1$ and $\kappa_2$ to satisfy $\kappa_1 \ll \kappa_2$, so that the visual effect as shown in Figures \ref{fig:sysm4} and \ref{fig:sysm5} is clearly manifested.

\bibliographystyle{ieee}
\bibliography{./wheat-chaff}

\begin{thebibliography}{1}

\bibitem{fcc10:order}
Federal Communication Commission,
\newblock ``Second memorandum opinion and order in the matter of unlicensed operation in the TV broadcast bands, additional spectrum for unlicensed devices below 900 MHz and in the 3 GHz band,''
\newblock {\it Docket Number 10-174}, Sep. 2010

\bibitem{cept11:requirements}
European Conference of Postal and Telecommunications Administrations,
\newblock ``Technical and operational requirements for the possible operation of cognitive radio systems in the `white spaces' of the frequency band 470-790 MHz,''
\newblock {\it ECC Report 159}, Cardiff, Jan. 2011

\bibitem{shellhammer11:chapter}
S.~J. Shellhammer, C. Shen, A.~K. Sadek and W. Zhang,
\newblock ``TV white space regulations,''
\newblock in \textit{Wireless Networks in the TV White Space: Concepts, Techniques and Applications}, R.~A. Saeed and S.~J. Shellhammer ed., CRC Press, 2011

\bibitem{fcc47cfr}
Federal Communication Commission,
\newblock ``Low power auxiliary stations,''
\newblock {\it FCC 47 CFR Ch. I (10-1-07 edition), Part 74, Subpart H}

\bibitem{oet08:report}
Technical Research Branch, Laboratory Division, Office of Engineering and Technology, Federal Communications Commission,
\newblock ``Evaluation of the performance of prototype TV-band white space devices phase II,''
\newblock {\it FCC/OET 08-TR-1005}, Oct. 2008

\bibitem{ghosh08crowncom}
M. Ghosh, V. Gaddam, G. Turkenich, and K. Challapali,
\newblock ``Spectrum sensing prototype for sensing ATSC and wireless microphone signals,''
\newblock in \textit{Proc. Int. Conf. Cognitive Radio Oriented Wireless Networks and Communications (CrownCom)}, 2008

\bibitem{shellhammer09:ita}
S.~J. Shellhammer, A.~K. Sadek, and W. Zhang,
\newblock ``Technical challenges for cognitive radio in the TV white space spectrum,''
\newblock in \textit{Proc. Inform. Theory and App. (ITA) Workshop}, San Diego, CA, USA, Feb. 2009

\bibitem{balamurthi11:dyspan}
R. Balamurthi, H. Joshi, C. Nguyen, A.~K. Sadek, S.~J. Shellhammer, and C. Shen,
\newblock ``A TV white space spectrum sensing prototype,''
\newblock in \textit{Proc. IEEE Int. Symp. Dynamic Spectrum Access Networks (DySPAN)}, Aachen, Germany, Apr. 2011

\bibitem{han06:icact}
N. Han, S. Shon, J.~H. Chung, and J.~M. Kim,
\newblock ``Spectral correlation based signal detection method for spectrum sensing in IEEE 802.22 WRAN systems,''
\newblock in \textit{Proc. 8th Int. Conf. Adv. Commun. Tech. (ICACT)}, Phoenix Park, Korea, Feb. 2006

\bibitem{kim07:dyspan}
K. Kim, I.~A. Akbar, K.~K. Bae, J. Urn, C.~M. Spooner, and J.~H. Reed,
\newblock ``Cyclostationary approaches to signal detection and classification in cognitive radio'',
\newblock in \textit{Proc. IEEE Int. Symp. Dynamic Spectrum Access Networks (DySPAN)}, Dublin, Ireland, Apr. 2007

\bibitem{zeng09:com}
Y. Zeng and Y.~C. Liang,
\newblock ``Eigenvalue-based spectrum sensing algorithms for cognitive radio,''
\newblock \textit{IEEE Trans. Commun.}, Vol. 57, No. 6, pp. 1784-1793, Jun. 2009

\bibitem{chen10:dyspan}
H.-S. Chen and W. Gao,
\newblock ``Spectrum sensing for FM wireless microphone signals,''
\newblock in \textit{Proc. IEEE Int. Symp. Dynamic Spectrum Access Networks (DySPAN)}, Singapore, Apr. 2010

\bibitem{gautier10:crowncom}
M. Gautier, M. Laugeois, and D. Noguet,
\newblock ``Teager-Kaiser energy detector for narrowband wireless microphone spectrum sensing,''
\newblock in \textit{Proc. Int. Conf. Cognitive Radio Oriented Wireless Networks and Communications (CrownCom)}, 2010

\bibitem{adoum10:iccce}
B.~A. Adoum and V. Jeoti,
\newblock ``Cyclostationary feature based multiresolution spectrum sensing approach for DVB-T and wireless microphone signals,''
\newblock in \textit{Proc. Int. Conf. Comput. Commun. Eng. (ICCCE)}, Kuala Lumpur, Malaysia, May 2010

\bibitem{zhang10:globecom}
D. Zhang, L. Dong, and N. Mandayam,
\newblock ``Sensing wireless microphone with ESPRIT from noise and adjacent channel interference,''
\newblock in \textit{Proc. IEEE Global Commun. Conf. (Globecom)}, Miami, FL, USA, Dec. 2010

\bibitem{802fcc}
IEEE P802.22 Wireless RANs,
\newblock ``FCC R\&O conference call minutes,''
\newblock {\it IEEE 802.22-08/0344r0}, Dec. 2008

\bibitem{sharkey09:petition}
S.~B. Sharkey and R.~D. Kubik,
\newblock ``Petition for reconsideration and clarification, Motorola, Inc.,''
\newblock Motorola, Inc., Mar. 2009

\bibitem{kailath98:it}
T. Kailath and H.~V. Poor,
\newblock ``Detection of stochastic processes,''
\newblock {\it IEEE Trans. Inform. Theory}, Vol. 44, No. 6, pp. 2230-2259, Oct. 1998

\bibitem{kullback59:book}
S. Kullback,
\newblock \textit{Information Theory and Statistics},
\newblock John Wiley \& Sons, New York, 1959

\bibitem{gardner86:assp}
W.~A. Gardner,
\newblock ``Measurement of spectral correlation,''
\newblock \textit{IEEE Trans. Acoust., Speech, and Signal Processing}, Vol. 34, No. 5, pp. 1111-1123, Oct. 1986

\bibitem{gardner86:sp}
W.~A. Gardner,
\newblock ``The spectral correlation theory of cyclostationary time-seris,''
\newblock \textit{Signal Processing}, Vol. 11, No. 1, pp. 13-36, Jul. 1986

\bibitem{gardner94:book}
W.~A. Gardner {\it et. al.},
\newblock \textit{Cyclostationarity in Communications and Signal Processing},
\newblock W.~A. Gardner ed., IEEE Press, 1994

\bibitem{sullivan95:sp}
M.~C. Sullivan and E.~J. Wegman,
\newblock ``Estimating spectral correlations with simple nonlinear transformations,''
\newblock \textit{IEEE Trans. Signal Processing}, Vol. 43, No. 6, pp. 1525-1526, Jun. 1995

\end{thebibliography}

\balance

\end{document}